\documentclass[aps,prb,twocolumn]{revtex4}
\usepackage{graphicx}

\usepackage{amssymb}
\usepackage{graphicx}
\usepackage{amsmath}
\usepackage{graphicx}
\usepackage{dcolumn}
\usepackage{bm}
\usepackage{color}

\begin{document}

\title{Tight-binding model and direct-gap/indirect-gap transition
in single-layer and multi-layer MoS$_2$}

\author{E. Cappelluti$^{1,2}$, R. Rold\'an$^2$, J.A. Silva-Guill\'en$^3$, P. Ordej\'on$^3$, and
  F. Guinea$^1$}

\affiliation{\centerline{$^1$Instituto de Ciencia de Materiales de Madrid,
CSIC, c/ Sor Juana Ines de la Cruz 3, 28049 Cantoblanco, Madrid, Spain} \\
\centerline{$^2$Istituto de Sistemi Complessi, U.O.S. Sapienza, CNR, v. dei
Taurini 19, 00185 Roma, Italy}\\
\centerline{$^3$Centre d'Investigaci\'o en Nanoci\`encia
i Nanotecnologia - CIN2 (CSIC-ICN), Campus UAB, Bellaterra, Spain}
}

\date{\today}

\begin{abstract}
In this paper we present a paradigmatic tight-binding model
for single-layer as well as for multi-layered semiconducting MoS$_2$ and similar
transition metal dichalcogenides. We show that the electronic
properties of multilayer systems can be reproduced
in terms of a tight-binding modelling of the single-layer
hopping terms  by simply adding the proper interlayer hoppings
ruled by the chalcogenide atoms.
We show that such tight-binding model permits to understand and
control in a natural way the transition between a direct-gap
band structure, in single-layer systems, to an indirect gap
in multilayer compounds in terms of a momentum/orbital selective
interlayer splitting of the relevant valence and conduction bands.
The model represents also a suitable playground to investigate
in an analytical way strain and finite-size effects.
\end{abstract}

\maketitle

\section{Introduction}

The isolation of flakes of single-layer and few-layer
graphene\cite{novoselov1,novoselov2,ZK05}
has triggered a huge burst of interest on
two-dimensional layered materials because of their structural
and electronic properties.
Due to its huge electronic mobility, graphene has been in the last years the
main focus of the research in the field.
However, a drawback in engineering graphene-based electronic
device is the absence of a gap in the monolayer samples,
and the difficulty in opening a gap in multilayer systems
without affecting the mobility.
As an alternative route, recent research is exploring the
idea of multilayered heterostructures built up from interfacing
different twodimensional materials.\cite{britnell}
Along this perspective, semiconducting dichalcogenides such as MoS$_2$, MoSe$_2$, 
WS$_2$, etc. are promising compounds since they
can be easily exfoliated and present a suitable small gap
both in single-layer and in few-layer samples.
Quite interestingly, in few-layer MoS$_2$ the size and the nature of the gap
depends on the number $N$ of MoS$_2$ layers, with a transition between
a direct gap in monolayer ($N=1$) compounds to a smaller indirect gap
for $N \ge 2$. \cite{li,lebegue,mak,splendiani}
In addition, the electronic properties appear to be highly sensitive
to the external pressure and
strain, which affect the insulating gap and, under particular conditions,
can also induce a insulator/metal transition. \cite{feng,lu,pan,peelaers,yun,scalise1,scalise2,li3,
ghorbani,shi,hromadova}
Another intriguing feature of these materials is the strong
entanglement between the spin and the orbital/valley degrees of
freedom, which permits, for instance, to manipulate spins by means of
circularly polarized light.\cite{mak12,mak13,cao,sallen,xiao,wu,zeng,OR13}
Moreover, in MoSe$_2$, a transition between a direct to an indirect gap
was observed as a function of temperature.\cite{tongay}

On the theoretical level, the description of its low-energy electronic
properties is enormously facilitated by the 
the availability of a paradigmatic Hamiltonian model
for the single-layer in terms of few tight-binding (TB)
parameters\cite{wallace,reich2002} (actually only one, the nearest neighbors 
carbon-carbon hopping $\gamma_0$, in the simplest case).\cite{pacoreview} 
The well-known Dirac equation can thus be derived
from that as a low-energy expansion.
Crucial to the development of the theoretical analysis
in graphene is also the fact that model Hamiltonians for multilayer
graphenes can be built using the single-layer TB
description as a fundamental block and just adding additional
interlayer hopping
terms.\cite{mccann,nilsson,pp1,pp2,abp,koshino1,koshino2,koshino3,
zhang,yuan,yan,olsen,mak4,lui,bao,ubrig}
Different stacking orders can be also easily investigated.
The advantage of such tight-binding description with respect
to first-principles calculations is that it provides a simple starting point
for the further inclusion of many-body electron-electron effects
by means of Quantum Field Theory (QFT)
techniques, as well as of the dynamical effects of the
electron-lattice interaction.
Tight-binding approaches can be also more convenient
than first-principles methods such as Density Functional Theory (DFT)
for investigating systems involving a very large number of atoms.
Although DFT methods are currently able to handle systems with hundreds
or even thousands of atoms\cite{siesta1,siesta2}, and have been thoroughly applied
to large scale graphene-related problems\cite{fuhrer,wang,lehtinen,novaes}, 
they are still computationally
challenging and demanding. Therefore, TB has been the method of choice for
the study of  disordered and inhomogeneous 
systems\cite{ho,palacios,pereira,carpio,lopezsancho,ribeiro,yuan1,yuan2,neek,yuan3,leconte2010,soriano2011,leconte2011,vantuan}
materials nanostructured in large scales (nanoribbons,
ripples)\cite{zheng,guinea08,castro1,castro2,castro3,castro4,cresti,huang,klymenko}
or in twisted multilayer materials.\cite{LdS,mele1,mele2,shallcross,morell,degail,mac,morell2,moon1,moon2,sanjose,tabert}

While much of the theoretical work of graphenic materials
has been based on tight-binding-like approaches,
the electronic properties of single-layer and few-layer dichalcogenides have been
so far mainly investigated by means of DFT
calculations.\cite{li,lebegue,splendiani,feng,lu,pan,peelaers,yun,scalise1,scalise2,li3,
ghorbani,shi,hromadova,kaasbjerg,kaasbjerg2,feng2,kadantsev,kosmider,zibouche,li4},
despite early work in non-orthogonal tight binding models for transition metal 
dichalcogenides.\cite{bromley}
Few simplified low-energy Hamiltonian models
has been presented for these materials, whose validity
is however restricted to the specific case of single-layer systems.
An effective low-energy model was for instance introduced
in Refs. \onlinecite{xiao,korma} to discuss the spin/orbital/valley coupling
at the K point. Being limited to the vicinity of the K point,
such model cannot be easily generalized to the multilayer case
where the gap is indirect with valence and conduction edges located
far from the K point.
An effective lattice TB Hamiltonian was on the other hand proposed
in Refs. \onlinecite{asgari}, valid in principle in the whole Brillouin
zone. However, the band structure of the
single-layer lacks the characteristic second minimum
in the conduction band (see later discussion) that will become
the effective conduction edge in multilayer systems,
so that also in this case the generalization to the multilayer
compounds is doubtful. In addition, the use of an overlap matrix
makes the proposed Hamiltonian unsuitable for a straightforward
use as a basis for QFT analyses.
This is also the case for a recent model proposed
in Ref. \onlinecite{zahid}, where
the large number (ninetysix) of free fitting parameters
and the presence of overlap matrix make
such model inappropriate for practical use
within the context of Quantum Field Theory.

In this paper we present a suitable tight-binding model
for the dichalcogenides
valid both in the single-layer case and in the multilayer one.
Using a Slater-Koster approach,\cite{sl} and focusing on MoS$_2$
as a representative case, we analyze the orbital character of the electronic states at the
relevant high-symmetry points.
Within this context we show that
the transition from a direct gap to an indirect gap
in MoS$_2$ as a function of the number of layers can be understood
and reproduced
in a natural way as a consequence of a momentum/orbital selective interlayer
splitting of the main relevant energy levels.
In particular, we show that the $p_z$ orbital of the S atoms
plays a pivotal role in such transition and it cannot be neglected
in reliable tight-binding models aimed to describe single-layer
as well as multi-layer systems.
The tight-binding description here introduced can represent thus
the paradigmatic model for the analysis of the electronic properties
in multilayer systems in terms of intra-layer ligands
plus  a finite number of interlayer hopping terms.
Such tight-binding model, within the context
of the Slater-Koster approach, provides also a suitable tool
to include in an analytical and intuitive way effects
of pressure/strain by means of the modulation of the interatomic distances.
The present analysis defines, in addition, the minimum constraints
that the model has to fulfill to guarantee a correct description
of the band structure of multi-layer compounds.

The paper is structured as follows:
in Section \ref{s:dft} we present DFT calculations for single-layer
and multi-layer (bulk) MoS$_2$, which will be here used as a reference for the
construction of a tight-binding model.
In Section \ref{s:1L} we describe the minimum tight-binding model
for the single-layer case needed to reproduce the fundamental
electronic properties and the necessary orbital content.
The decomposition of the Hamiltonian in blocks
and the specific orbital character at the high-symmetry points is
discussed. The extension of the tight-binding model to the bulk case,
taken as representative of multilayer compounds, is addressed in
Section \ref{s:bulk}. We pay special attention to reveal the microscopic origin of the change
between a direct-gap to indirect-gap band structure.
In Section \ref{s:disc} we summarize the implications of our analysis
in the building of a reliable tight-binding model, and we
provide a possible set of tight-binding parameters
for the single-layer and multilayer case.


\section{DFT calculations and orbital character}
\label{s:dft}

\begin{figure}[t]
\includegraphics[scale=0.37,clip=]{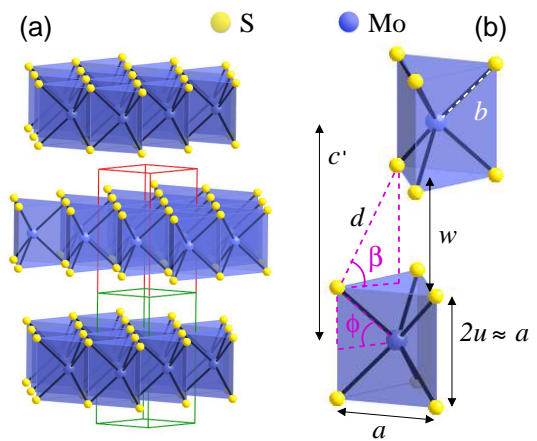}
\caption{(a) Model of the atomic structure of MoS$_2$.
The bulk compound has a 2H-MoS$_2$ structure
with two MoS$_2$ layers per unit cell, each layer 
being built up from a trigonal prism coordination unit.
The small green rectangle represents the unit cell of a monolayer of MoS$_2$,
which is doubled (red extension) in the bulk crystal. 
(b) Detail of the trigonal prisms for the two layers in the bulk compound,
showing the lattice constants and the definition
of the structural angles used in the text.
}
\label{f-structure}
\end{figure}

In the construction of a reliable TB model for semiconducting
dichalcogenides we will be guided by first-principles DFT calculations
taht will provide the reference on which to calibrate the TB
model.
We will focus here on MoS$_2$ as a representative case,
although we have performed first-principle calculations
for comparison also on WS$_2$.
The differences in the electronic structure and in the
orbital character of these two compounds are, however, minimal
and they do not involve any different physics.
The structure of single-layer and multilayer MoS$_2$
is depicted in Fig. \ref{f-structure}.

The basic unit block is composed of an inner layer of Mo atoms
on a triangular lattice sandwiched between two layers
of S atoms lying on the triangular net of alternating
hollow sites. Following standard notations,\cite{bromley}
we denote $a$ as the distance between
nearest neighbor in-plane Mo-Mo and S-S distances,
$b$ as the nearest neighbor Mo-S distance and
$u$ as the distance between the Mo and S planes.
The MoS$_2$ crystal forms an almost perfect trigonal prism
structure with $b$ and $u$ very close to the their ideal values 
$b \simeq \sqrt{7/12}a$ and $u \simeq a/2$.
In our DFT calculations, we use experimental values 
for bulk MoS$_2$,\cite{bromley}
namely $a=3.16$ \AA, $u=1.586$ \AA,
and, in bulk systems, a distance between Mo planes as $c'=6.14$ \AA,
with a lattice constant in the 2H-MoS$_2$ structure of $c=2c'$.
The in-plane Brillouin zone is thus characterized by the
high-symmetry points $\Gamma=(0,0)$, K$=4\pi/3a(1,0)$,
and M$=4\pi/3a (0,\sqrt{3}/2)$.
DFT calculations are done using
the \textsc{Siesta} code.\cite{siesta1,siesta2}
We use the exchange-correlation potential of Ceperly-Alder\cite{ceperley_alder_1980}
as parametrized by Perdew and Zunger.\cite{perdew}
We use also a split-valence double-$\zeta $ basis set including
polarization functions.\cite{arsan99}
The energy cutoff and the Brillouin zone sampling were chosen to converge the total energy.

\begin{figure}[t]
\includegraphics[scale=0.4,clip=]{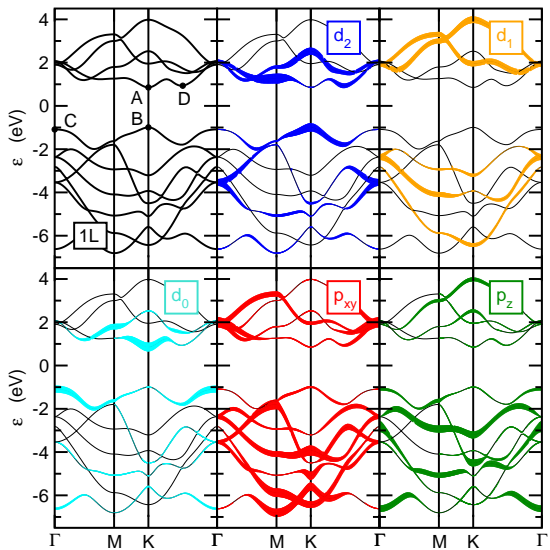}
\caption{Band structure and orbital character of single-layer MoS$_2$.
The top left panel shows the
full band structure while, in the other panels, the thickness
of the bands represents the orbital weight, where the $d$-character
($d_2=d_{x^2-y^2}$, $d_{xy}$,
$d_1=d_{xz}$, $d_{yz}$,
$d_0=d_{3z^2-r^2}$)
refers to the Mo atom 4$d$ orbitals, while
the $p$-character
($p_{xy}=p_x$, $p_y$) refers to 2$p$ orbitals of sulfur.
}
\label{f-bandsDFTmos2}
\end{figure}

The electronic dispersion for the single-layer MoS$_2$ is nowadays well known.
We will only focus on the block of bands  containing the first four conduction
bands and by the first seven valence bands,
in an energy window of from -7 to 5 eV around the Fermi level.
Our DFT calculations are shown in Fig. \ref{f-bandsDFTmos2},
where we show the orbital character of each band.
%
%
We use here the shorthand notation
$d_2$ to denote Mo $4d_{x^2-y^2}$, $4d_{xy}$orbitals;
$d_1$ for the Mo $4d_{xz}$, $4d_{yz}$ orbitals;
$d_0$ for the Mo $4d_{3z^2-r^2}$ orbital;
$p_{xy}$ (or simply $p$) to denote the S $3p_x$, $3p_y$ orbitals;
and
$p_z$ (or simply $z$) for the S $3p_z$orbital.
The four conduction bands and the seven valence bands
are mainly constituted by the five 4$d$ orbitals of Mo and
the six (three for each layer) 3$p$ orbitals of S, which
sum up to the $93$ \% of the total orbital weight
of these bands.

A special role in the electronic properties of these materials
is played by the electronic states labeled as (A)-(D)
and marked with black bullets
in Fig.  \ref{f-bandsDFTmos2}.
A detailed analysis of the orbital character of each energy level
at the main high-symmetry points of the Brillouin zone,
as calculated by DFT, is provided in Table \ref{t-block}.
\begin{table}
\begin{tabular}{cccccc}
\hline
\hline
energy    & main  & second & other & sym. & TB \\
DFT (eV)  & orb. &  orb.  & orbs. &     & label \\
\hline
\multicolumn{6}{c}{$\Gamma$ point} \\
\hline
2.0860$^*$ & 68 \% $p_{x/y}$ & 29 \% $d_2$ & 3 \% & E &
$E_{pd_2,+}(\Gamma)$ \\
1.9432$^*$  & 58 \% $p_{x/y}$ & 36 \% $d_1$ & 6 \% & O &
$E_{pd_1,+}(\Gamma)$ \\
 -1.0341 & 66 \% $d_0$ & 28 \% $p_z$
& 6 \% & E & $E_{zd_0,+}(\Gamma)$ \\
-2.3300$^*$ &  54 \% $d_1$ & 42 \% $p_{x/y}$
& 4 \% & O & $E_{pd_1,-}(\Gamma)$ \\
-2.6801  &100 \% $p_z$ & -  & 0 \% & O &
$E_z(\Gamma)$ \\
-3.4869$^*$ & 65 \% $d_2$ & 32 \% $p_{x/y}$
& 3 \% & E & $E_{pd_2,-}(\Gamma)$ \\
-6.5967 & 57 \% $p_z$ & 23 \% $d_0$
& 20 \% & E & $E_{zd_0,-}(\Gamma)$ \\
\hline
\multicolumn{6}{c}{K point} \\
\hline
4.0127  & 60 \% $d_1$ & 36 \% $p_z$
& 4 \% & O & $E_{zd_1,+}(K)$ \\
 2.5269  &  65 \% $d_2$ & 29 \% $p_z$
& 6 \% & E & $E_{zd_2,+}(K)$ \\
1.9891 & 50 \% $d_1$ & 31 \% $p_{x/y}$
& 19 \% & O & $E_{pd_1,+}(K)$ \\
0.8162 & 82 \% $d_0$ & 12 \% $p_{x/y}$
& 6 \% & E & $E_{pd_0,+}(K)$ \\
-0.9919 & 76 \% $d_2$ & 20 \% $p_{x/y}$
& 4 \% & E & $E_{pd_2,+}(K)$ \\
-3.1975 & 67 \% $p_z$ & 27 \% $d_1$ 
& 6 \% & O & $E_{zd_1,-}(K)$ \\
-3.9056  & 85 \% $p_{x/y}$ & - 
& 15 \% & O & $E_{p}(K)$ \\
-4.5021  & 65 \% $p_z$ & 25 \% $d_2$
& 10 \% & E & $E_{zd_2,-}(K)$ \\
-5.0782  & 71 \% $p_{x/y}$ & 12 \% $d_2$ 
& 17 \% & E & $E_{pd_2,-}(K)$ \\
-5.5986  & 66 \% $p_{x/y}$ & 14 \% $d_0$
& 20 \% & E & $E_{pd_0,-}(K)$ \\
 -6.4158 & 60 \% $p_{x/y}$ & 37 \% $d_1$
& 3 \% & O & $E_{pd_1,-}(K)$ \\
\hline
\hline
\end{tabular}
\mbox{} $^*$Double-degenerate level \hfill \mbox{}
\caption{Energy levels and orbital content
of single-layer MoS$_2$
evaluated by DFT calculations.
We report here the first two main orbital characters belonging
to the blocks Mo-$4d$ and S-$3p$, while the following column
shows the remaining character not belonging to these orbital group.
Also show is the association
of each level with the corresponding eigenvalue
of the tight-binding model and the
symmetry with respect
to the $z \rightarrow -z$ inversion
(E=even, O=odd).
The label $E_{\alpha\beta,\pm}$ 
in the last column
denotes the orbital character of the TB eigenstate,
with $\alpha,\beta=p,z,d_2,d_1,d_0$,
where $p=p_x,p_y$, $z=p_z$, $d_2=d_{x^2-y^2}, d_{xy}$,
$d_1=d_{xz}, d_{yz}$, $d_0=d_{3z^2-r^2}$.
The index $\pm$ denotes the higher energy [$(+)= $ antibonding]
and the lower energy [$(-)= $ bonding].
}
\label{t-block}
\end{table}
We can notice that
an accurate description
of the conduction and valence band edges (A)-(B) at the K point
involves at least the Mo orbitals $d_{3z^2-r^2}$, $d_{x^2-y^2}$, $d_{xy}$,
and the S orbitals $p_{x}$, $p_y$.
Along this perspective, a 5-band tight-binding model,
restricted to the subset of these orbitals, was presented in
Ref. \onlinecite{asgari}, whereas
even the S $3p$ orbitals were furthermore omitted in
Ref. \onlinecite{xiao}.

The failure of this latter orbital restriction for a more comprehensive
description is however pointed out when analyzing other relevant
high-symmetry Brillouin points.
In particular, concerning the valence band, we can notice
a second maximum at the $\Gamma$ point, labeled as (C)
in Fig.  \ref{f-bandsDFTmos2},
just $42$ meV below the real band edge at the K point
and with main $d_0$-$p_z$ orbital character.
The relevance of this secondary band extreme is evident
in the multilayer compounds ($N \ge 2$), where such maximum
at $\Gamma$ increases its energy to become the
effective band edge. \cite{li,splendiani}

\begin{figure}[t]
\includegraphics[scale=0.4,clip=]{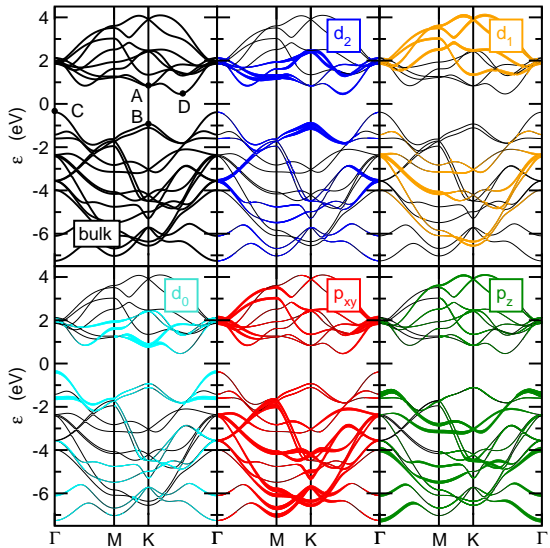}
\caption{Band structure and orbital character for 
bulk 2H-MoS$_2$.
Labels similar as in Fig. \ref{f-bandsDFTmos2}.
}
\label{f-bandsDFTmos2bulk}
\end{figure}

The band structure with the orbital character for the bulk
($N=\infty$) case, representative of the multilayer case, is shown
in Fig. \ref{f-bandsDFTmos2bulk}.
A similar change of the topology of the band edge occurs
in the conduction band.
Here a secondary minimum, labeled as (D)
in Fig.  \ref{f-bandsDFTmos2}, at $Q=4\pi/3a(1/2,0)$,
midway along the $\Gamma$-K cut, is present in the single-layer
compounds.
Such minimum however moves down in energy
in multilayer systems to become the effective conduction
band edge.\cite{li,splendiani}
Even in this case, a relevant $p_z$ component is involved
in the orbital character of this electronic state.
The topological changes of the location of the band edges
in the Brillouin zone are responsible for the observed switch
from a direct to an indirect gap in multilayer samples.
As we will see, thus, the inclusion of the $p_z$ orbitals
in the full tight-binding Hamiltonian is not only desirable
for a more complete description, but it is also {\em unavoidable}
to understand the evolution of the band structure
as a function of the number of layers.

\section{Tight-binding description of the single-layer}
\label{s:1L}

The aim of this section is to define a tight-binding model for the
single-layer which will be straightforwardly generalizable
to the multilayer case by adding the appropriate interlayer hopping.
We will show that, to this purpose, all the $4d$ Mo orbitals and
the $3p$ S orbitals are needed to be taken into account.
Considering that the unit cell contains two S atoms,
we define the Hilbert space by means of the
11-fold vector:
\begin{widetext}
\begin{eqnarray}
\phi_i^\dagger
&=&
(
p_{i,x,t}^\dagger,
p_{i,y,t}^\dagger,
p_{i,z,t}^\dagger,
d_{i,3z^2-r^2}^\dagger,
d_{i,x^2-y^2}^\dagger,
d_{i,xy}^\dagger,
d_{i,xz}^\dagger,
d_{i,yz}^\dagger,
p_{i,x,b}^\dagger,
p_{i,y,b}^\dagger,
p_{i,z,b}^\dagger
),
\label{basis}
\end{eqnarray}
\end{widetext}
where $d_{i,\alpha}$ creates an electron 
in the orbital $\alpha$ of the Mo atom in the $i$-unit cell,
$p_{i,\alpha,t}$ creates an electron 
in the orbital $\alpha$ of the top ($t$) layer atom S in the $i$-unit cell,
and $p_{i,\alpha,b}$ creates an electron 
in the orbital $\alpha$ of the bottom ($b$) layer atom S in the $i$-unit
cell.

Once the Hilbert space has been introduced, the tight-binding model
is defined by the hopping integrals between the different orbitals,
described, in the framework of a Slater-Koster description,
in terms of $\sigma$, $\pi$ and
$\delta$ ligands.\cite{sl}
In order to provide a tight-binding model as a suitable basis for
the inclusion of many-body effects by means of diagrammatic techniques,
we assume that the basis orbitals are orthonormal, so that the overlap matrix is the unit matrix.
A preliminary analysis based on the interatomic distance can be useful
to identify the most relevant hopping processes.
In particular,
these are expected to be the ones
between nearest neighbor Mo-S
(interatomic distances $b=2.41$ \AA)
and between the nearest neighbor in-plane Mo-Mo and between the nearest neighbor in-plane and out-of-plane S-S atoms
(interatomic distance $a=3.16$ \AA).
Further distant atomic bonds,
in single-layer systems,
start from hopping between second nearest
neighbor Mo-S atoms, with interatomic distance
$3.98$ \AA, and they will be here discarded.

All the hopping processes of the relevant pair of neighbors
are described in terms
of the Slater-Koster parameters, respectively
$V_{pd\sigma}$, $V_{pd\pi}$ (Mo-S bonds),
$V_{dd\sigma}$, $V_{dd\pi}$, $V_{dd\delta}$ (Mo-Mo bonds),
and $V_{pp\sigma}$, $V_{pp\pi}$ (S-S bonds).
Additional relevant parameters are the crystal fields
$\Delta_0$, $\Delta_1$,  $\Delta_2$, $\Delta_p$, $\Delta_z$, 
describing respectively the atomic level the $l=0$ ($d_{3z^2-r^2}$),
 the $l=1$ ($d_{xz}$, $d_{yz}$),
 the $l=2$ ($d_{x^2-y^2}$, $d_{xy}$) Mo orbitals,
 the in-plane ($p_x$, $p_y$) S orbitals
and of the out-of-plane $p_z$ S orbitals.
We end up with a total of 12 tight-binding parameters to be determined,
namely: $\Delta_0$, $\Delta_1$,  $\Delta_2$, $\Delta_p$, $\Delta_z$, 
$V_{dd\sigma}$, $V_{dd\pi}$, $V_{dd\delta}$, $V_{pp\sigma}$,
$V_{pp\pi}$, $V_{pd\sigma}$, $V_{pd\pi}$.

In the orbital basis of Eq. (\ref{basis}), we
can write thus the tight-binding Hamiltonian in the form:
\begin{eqnarray}
H
&=&
\sum_{\bf k}
\phi_{\bf k}^\dagger
\hat{H}_{\bf k}
\phi_{\bf k},
\label{ham}
\end{eqnarray}
where $\phi_{\bf k}$ is the Fourier transform of $\phi_i$
in  momentum space.
The Hamiltonian matrix
can be written (we drop for simplicity from now on the index ${\bf k}$)
as:
\begin{eqnarray}
\hat{H}
&=&
\left(
\begin{array}{ccc}
\hat{H}_{pt,pt} & \hat{H}_{d,pt}^\dagger & \hat{H}_{pt,pb} \\
\hat{H}_{d,pt} & \hat{H}_{d,d} & \hat{H}_{d,pb} \\
\hat{H}_{pb,pb}^* & \hat{H}_{d,pb}^\dagger &\hat{H}_{pb,pb} 
\end{array}
\right),
\label{hmatr11}
\end{eqnarray}
where $\hat{H}_{pb,pb}=\hat{H}_{pt,pt}$ describes the in-plane hopping
in the top and bottom S layer, namely,
\begin{eqnarray}
\hat{H}_{pb,pb}
&=&
\hat{H}_{pt,pt}
=
\left(
\begin{array}{ccc}
H_{x/x} & H_{x/y} & 0 \\
H_{x/y}^* & H_{y/y} & 0 \\
0 & 0 & H_{z/z} 
\end{array}
\right),
\label{hmatrpp}
\end{eqnarray}
$\hat{H}_{d,d}$ the in-plane hopping in the middle Mo layer, namely,
\begin{eqnarray}
\hat{H}_{d,d}
&=&
\left(
\begin{array}{ccccc}
H_{z^2/z^2} & H_{z^2/x^2} & H_{z^2/xy} & 0 & 0 \\
H_{z^2/x^2}^* & H_{x^2/x^2} & H_{x^2/xy} & 0 & 0 \\
H_{z^2/xy}^* & H_{x^2/xy}^* & H_{xy/xy} & 0 & 0 \\
0 & 0 & 0 & H_{xz/xz} & H_{xz/yz} \\
0 & 0 & 0 & H_{xz/yz}^* & H_{yz/yz} 
\end{array}
\right),
\label{hmatrdd}
\end{eqnarray}
$\hat{H}_{pt,pb}$ the vertical hopping between S orbitals
in the top and bottom layer,
\begin{eqnarray}
\hat{H}_{pt,pb}
&=&
\left(
\begin{array}{cccc}
V_{pp\pi} & 0 & 0 \\
0 & V_{pp\pi} & 0 \\
0 & 0 & V_{pp\sigma} 
\end{array}
\right),
\label{hmatrptpb}
\end{eqnarray}
and $\hat{H}_{d,pt}$, $\hat{H}_{d,pb}$ the hopping
between Mo and S atoms in the top and bottom planes,
respectively:
\begin{eqnarray}
\hat{H}_{d,pt}
&=&
\left(
\begin{array}{ccc}
H_{z^2/x} & H_{z^2/y} & H_{z^2/z} \\
H_{x^2/x} & H_{x^2/y} & H_{x^2/z} \\
H_{xy/x} & H_{xy/y} & H_{xy/z} \\
H_{xz/x} & H_{xz/y} & H_{xz/z}\\
H_{yz/x} & H_{yz/y} & H_{yz/x} 
\end{array}
\right),
\label{hmatrdpt}
\end{eqnarray}
\begin{eqnarray}
\hat{H}_{d,pt}
&=&
\left(
\begin{array}{ccc}
H_{z^2/x} & H_{z^2/y} & -H_{z^2/z} \\
H_{x^2/x} & H_{x^2/y} & -H_{x^2/z} \\
H_{xy/x} & H_{xy/y} & -H_{xy/z} \\
-H_{xz/x} & -H_{xz/y} & H_{xz/z}\\
-H_{yz/x} & -H_{yz/y} & H_{yz/x} 
\end{array}
\right).
\label{hmatrdpb}
\end{eqnarray}
Here and in the following, for  the sake of compactness,
we use the shorthand notation
$3z^2-r^2  \Rightarrow z^2$ and $x^2-y^2  \Rightarrow x^2$. An explicit expression for the different Hamiltonian matrix elements
in terms of the Slater-Koster tight-binding parameters
can be provided following the seminal work by Doran {\em et al.}
(Ref. \onlinecite{doran}) and it is reported for completeness in
Appendix \ref{app_elements}.

Eqs. (\ref{ham})-(\ref{hmatrdpb}) define our tight-binding model
in terms of a $11 \times 11$ Hamiltonian $\hat{H}$
which can be now explicitly solved to get
eigenvalues and eigenvectors in the whole Brillouin zone or along
the main axes of high symmetry.
It is now an appealing task to associate each DFT energy level
with the Hamiltonian eigenvalues, whose eigenvectors will shed
light on the properties of the electronic states.
Along this line, we are facilitated by 
symmetry arguments which permit,
in the monolayer compounds,
to decoupled the $11 \times 11$ Hamiltonian in Eq. (\ref{hmatr11}),
in two main blocks, with different symmetry with respect to
the mirror inversion $z \rightarrow -z$.\cite{doran}
This task is accomplished by introducing a symmetric and antisymmetric
linear combination of the $p$ orbital of the S atoms on the top/bottom
layers. 
More explicitly, we use the basis vector
\begin{widetext}
\begin{eqnarray}
\tilde{\phi}_k^\dagger
&=&
(
d_{k,3z^2-r^2}^\dagger,
d_{k,x^2-y^2}^\dagger,
d_{k,xy}^\dagger,
p_{k,x,S}^\dagger,
p_{k,y,S}^\dagger,
p_{k,z,A}^\dagger,
d_{k,xz}^\dagger,
d_{k,yz}^\dagger,
p_{k,x,A}^\dagger,
p_{k,y,A}^\dagger,
p_{k,z,S}^\dagger
),
\label{basistilde}
\end{eqnarray}
\end{widetext}
where
$p_{k,\alpha,S}^\dagger=(p_{k,\alpha,t}^\dagger+p_{k,\alpha,b}^\dagger)/\sqrt{2}$,
$p_{k,\alpha,A}^\dagger=(p_{k,\alpha,t}^\dagger-p_{k,\alpha,b}^\dagger)/\sqrt{2}$.
Note that our basis differs slightly with respect to the one 
employed in Ref. \onlinecite{doran} because we have introduced
explicitly the proper normalization factors to make it unitary.
In this basis we can write thus
\begin{eqnarray}
\hat{H}
&=&
\left(
\begin{array}{cc}
\hat{H}_{\rm E} & 0 \\
0 & \hat{H}_{\rm O}
\end{array}
\right),
\label{hdecoupled}
\end{eqnarray}
where $\hat{H}_{\rm E}$ is a $6 \times 6$ block
with even (E) symmetry with respect to the mirror
inversion $z \rightarrow -z$,
and $\hat{H}_{\rm O}$ a $5 \times 5$ block with odd (O)
symmetry.
We should remark however that such decoupling holds true
only in the single-layer case and only in the absence
of a $z$-axis electric field, as it can be induced by substrates or
under gating conditions.
In the construction of a tight-binding model that could permit
a direct generalization to the multilayer case, the interaction between the band blocks with even and odd symmetry
should be thus explicitly retained.

The association between DFT energy levels and tight-binding
eigenstates is now further simplified on specific
high-symmetry points of the Brillouin zone.
Most important are the $K$ and the $\Gamma$ points,
which are strictly associated with the direct and indirect gap
in monolayer and multilayered compounds.

\subsection{$\Gamma$ point}

We present here a detailed analysis of the eigenstates and  their orbital character at the $\Gamma$
point.
For the sake of simplicity, we discuss separately the blocks with even and odd 
 symmetry with respect to the inversion
$z \rightarrow -z$.
The identification of the DFT levels with the tight-binding eigenstates
is facilitated by the possibility of decomposing the full Hamiltonian
in smaller blocks, with typical size $2 \times 2$ (dimers)
or $1 \times 1$ (monomers).
In particular, the  $6 \times 6$ block with even symmetry can be
decomposed (see Appendix \ref{app_blocks} for details) as:
\begin{eqnarray}
\hat{H}_{\rm E}(\Gamma)
&=&
\left(
\begin{array}{ccc}
\hat{H}_{zd_0}(\Gamma) & 0 &  0 \\
0 & \hat{H}_{pd_2}(\Gamma) &  0 \\
0 & 0 & \hat{H}_{pd_2}(\Gamma)
\end{array}
\right).
\label{hgamma66}
\end{eqnarray}
Here each matrix, $\hat{H}_{pd_2}$, $\hat{H}_{zd_0}$ represents a
$2\times 2$ block where the indices describe the orbital character
of the dimer.
In particular, $\hat{H}_{pd_2}$
involves only $d_2=d_{x^2-y^2}, d_{xy}$ Mo-orbitals
and $p_x, p_y$ S-orbitals, whereas
$\hat{H}_{zd_0}$ involves only
the $d_0=d_{3z^2-r^2}$ Mo-orbital
and the $p_z$ S-orbital.
As it is evident in (\ref{hgamma66}),
the block $\hat{H}_{pd_2}$ appears twice and it is thus double
degenerate.
Similarly, we have
\begin{eqnarray}
\hat{H}_{\rm O}(\Gamma)
&=&
\left(
\begin{array}{ccc}
\hat{H}_{pd_1}(\Gamma) & 0 &  0 \\
0 & \hat{H}_{pd_1}(\Gamma) &  0 \\
0 & 0 & \Gamma_z^{\rm O}
\end{array}
\right),
\label{hgamma55}
\end{eqnarray}
where the doubly degenerate block $\hat{H}_{pd_1}$
involves only $d_1=d_{xz}, d_{yz}$ Mo-orbitals
and $p_x, p_y$ S-orbitals, while
$\Gamma_z^{\rm O}$ is a $1 \times 1$ block (monomer)
with pure character $p_z$.

It is also interesting to give a closer look at the inner structure
of a generic Hamiltonian sub-block.
Considering for instance $\hat{H}_{zd_0}$ as an example,
we can write
\begin{eqnarray}
\hat{H}_{zd_0}(\Gamma)
&=&
\left(
\begin{array}{cc}
\Gamma_0 & \sqrt{2}\Gamma_{zd_0}  \\
\sqrt{2}\Gamma_{zd_0} & \Gamma_z^{\rm E}
\end{array}
\right),
\end{eqnarray}
where $\Gamma_0$ is an energy level  with pure Mo $d_0$ orbital character
and $\Gamma_z^{\rm E}$ an energy level  with pure S $p_z$ orbital
character.
The off-diagonal term $\sqrt{2}\Gamma_{zd_0}$ acts thus here as a
``hybridization'', mixing the pure orbital character of $\Gamma_0$
and $\Gamma_z^{\rm E}$.
The suffix ``E''  here reminds that the level $\Gamma_z^{\rm E}$
belongs to the even symmetry block,
and it is useful to distinguish this state from a similar one
with odd symmetry (and different energy).
Keeping $\hat{H}_{zd_0}$ as an example,
the eigenvalues of a generic $2 \times 2$ block can be
obtained analytically:
\begin{eqnarray}
E_{zd_0,\pm}(\Gamma)
&=&
\frac{\Gamma_0+\Gamma_z^{\rm E}}{2}
\pm
\sqrt{\left(\frac{\Gamma_0-\Gamma_z^{\rm E}}{2}\right)^2+2\Gamma_{zd_0}^2}.
\end{eqnarray}

The explicit expressions of $\Gamma_\alpha$ and $\Gamma_{\alpha\beta}$
in terms of the Slater-Koster tight-binding parameters is reported
in Appendix \ref{app_elements}.\\
It is interesting to note that the diagonal
terms $\Gamma_\alpha$ ($\alpha=d_0,d_1,d_2,p,z$) are purely determined
by the crystal fields $\Delta_\alpha$ and by the tight-binding
parameters $V_{dd\sigma}$, $V_{dd\pi}$, $V_{dd\delta}$, 
$V_{pp\sigma}$, $V_{pp\pi}$, connecting Mo-Mo and S-S atoms, whereas
the hybridization off-diagonal terms $\Gamma_{\alpha\beta}$
depend exclusively on the Mo-S nearest neighbor hopping
$V_{pd\sigma}$,
$V_{pd\pi}$.

A careful comparison between the orbital character of each eigenvector
with the DFT results permits now to identify in an unambiguous way
each DFT energy level with its analytical tight-binding counterpart.
Such association is reported in Table \ref{t-block}, where
also the even/odd symmetry inversion is considered.

The use of the present analysis to characterize the properties
of the multilayer MoS$_2$ will be discussed in Section \ref{s:bulk}.

\subsection{K point}

A crucial role in the properties of semiconducting dichalcogenides is
played by the K point in the Brillouin zone, where the direct
semiconducting gap occurs in the single-layer systems.
The detailed analysis of the electronic spectrum is also favored here 
by the possibility of reducing the full $11 \times 11$ Hamiltonian in smaller
sub-blocks.
This feature is, however, less evident than at the $\Gamma$ point.
The even and odd components of the Hamiltonian take the form:
\begin{widetext}
\begin{eqnarray}
\hat{H}_{\rm E}(K)
&=&
\left(
\begin{array}{cccccc}
K_0 & 0 &  0 &  -i\sqrt{2}K_{pd_0} &\sqrt{2}K_{pd_0} & 0 \\
0 & K_2 &  0 &  i\sqrt{2}K_{pd_2} & \sqrt{2} K_{pd_2} & \sqrt{2}K_{zd_2} \\
0 & 0 &  K_2 &   -\sqrt{2} K_{pd_2} & i\sqrt{2} K_{pd_2}&
-i\sqrt{2}K_{zd_2} \\
i\sqrt{2}K_{pd_0} & -i\sqrt{2} K_{pd_2} &  -\sqrt{2} K_{pd_2} &
K_p^{\rm E} & 0 & 0 \\
\sqrt{2}K_{pd_0} & \sqrt{2} K_{pd_2} & -i\sqrt{2} K_{pd_2} &  0 &
K_p^{\rm E} & 0 \\
0 & \sqrt{2}K_{zd_2} &  i\sqrt{2}K_{zd_2} &  0 & 0 & K_z^{\rm E} \\
\end{array}
\right),
\label{hmatr66K}
\end{eqnarray}
\begin{eqnarray}
\hat{H}_{\rm O}(K)
&=&
\left(
\begin{array}{ccccc}
K_1 & 0 & \sqrt{2}K_{pd_1} & -i\sqrt{2}K_{pd_1} & -i\sqrt{2}K_{zd_1} \\
0 & K_1 & -i\sqrt{2}K_{pd_1} & -\sqrt{2}K_{pd_1} & \sqrt{2}K_{zd_1} \\
\sqrt{2}K_{pd_1} & i\sqrt{2}K_{pd_1} &  K_p^{\rm O} & 0 & 0 \\
i\sqrt{2}K_{pd_1} & -\sqrt{2}K_{pd_1} & 0 & K_p^{\rm O} & 0 \\
i\sqrt{2}K_{zd_1} & \sqrt{2}K_{zd_1} & 0 & 0 & K_z^{\rm O} 
\end{array}
\right).
\label{hmatr55k}
\end{eqnarray}
\end{widetext}
As for the $\Gamma$ point,
also here the upper labels ($\mu=$E, O) in $K_\alpha^\mu$ ($\mu=$E, O) express the
symmetry of the state corresponding to the energy level $K_\alpha^\mu$
with respect to the $z \rightarrow -z$ inversion.
The electronic properties of the Hamiltonian at the K point look
more transparent by introducing a different ``chiral'' base:
\begin{widetext}
\begin{eqnarray}
\bar{\psi}_{k}^\dagger
&=&
(
d_{k,3z^2-r^2}^\dagger,
d_{k,L2}^\dagger,
d_{k,R2}^\dagger,
p_{k,L,S}^\dagger,
p_{k,R,S}^\dagger,
p_{k,z,A}^\dagger,
d_{k,L1}^\dagger,
d_{k,R1}^\dagger,
p_{k,L,A}^\dagger,
p_{k,R,A}^\dagger,
p_{k,z,S}^\dagger
),
\label{psi}
\end{eqnarray}
\end{widetext}
where $d_{k,L2}=(d_{k,x^2-y^2}-id_{k,xy})/\sqrt{2}$,
$d_{k,R2}=(d_{k,x^2-y^2}+id_{k,xy})/\sqrt{2}$,
$d_{k,L1}=(d_{k,xz}-id_{k,yz})/\sqrt{2}$,
$d_{k,R1}=(d_{k,xz}+id_{k,yz})/\sqrt{2}$,
$p_{k,L,S}=(p_{k,x,S}-ip_{k,y,S})/\sqrt{2}$,
$p_{k,R,S}=(p_{k,x,S}+ip_{k,y,S})/\sqrt{2}$,
$p_{k,L,A}=(p_{k,x,A}-ip_{k,y,A})/\sqrt{2}$,
$p_{k,R,A}=(p_{k,x,A}+ip_{k,y,A})/\sqrt{2}$.

In this basis, the Hamiltonian matrix
can be also divided in smaller sub-blocks (see Appendix \ref{app_blocks}) as:
\begin{eqnarray}
\hat{H}_{\rm E}(K)
&=&
\left(
\begin{array}{ccc}
\hat{H}_{pd_0}(K) & 0 &  0 \\
0 & \hat{H}_{zd_2}(K) &  0 \\
0 & 0 & \hat{H}_{pd_2}(K) 
\end{array}
\right),
\label{hk66}
\end{eqnarray}
and
\begin{eqnarray}
\hat{H}_{\rm O}
&=&
\left(
\begin{array}{ccc}
\hat{H}_{pd_1}(K) & 0 &  0 \\
0 & \hat{H}_{zd_1}(K) &  0 \\
0 & 0 & K_{p}^{\rm O} 
\end{array}
\right).
\label{hk55}
\end{eqnarray}

As it is evident from the labels, each sub-block is also here
a $2 \times 2$ dimer, apart from the term $K_p^O$
which is a $1 \times 1$ block (monomer) with pure $p_x, p_y$
character.
The association between
the DFT energy levels and the tight-binding eigenstates
is reported also for the K point in Table \ref{t-block}.

\subsection{Q point}

As discussed above,
another special point determining the electronic properties
of MoS$_2$ is the Q point, halfway between the $\Gamma$ and K points
in the Brillouin zone, where the conduction band, in the single-layer
system, has a secondary minimum in addition to the absolute one
at the K point.
Unfortunately, not being a point of high-symmetry, the tight-binding Hamiltonian
cannot be decomposed in this case in simpler smaller blocks.
Each energy eigenvalue will contain thus a finite component of all
the Mo and S orbitals.
In particular, focusing on the secondary minimum in Q,
DFT calculations give 46 \% $d_2$, 24 \% $p_{x/y}$, 11 \% $p_z$ and 9 \% $d_0$.
The orbital content of this level will play a crucial role
in determining the band structure of multilayer compounds.

\subsection{Orbital constraints for a tight-binding model}

After having investigated in detail the orbital contents of each eigenstate
at the high-symmetry points, and having identified them
with the corresponding DFT energy levels, we can now employ such
analysis to assess the basilar conditions that a tight-binding model
must fulfill and to elucidate the physical consequences.

A first interesting issue is about the minimum number of orbitals
needed to be taken into account in a tight-binding model
for a robust description of the
electronic properties of these materials. A proper answer to such issue
is, of course, different
if referred to single-layer or multilayer compounds. For the moment we
will focus only on the single-layer case but we will underline on the
way the relevant features that will be needed to take into account
in multi-layer systems.

\begin{figure}[t]
\includegraphics[scale=0.35,clip=]{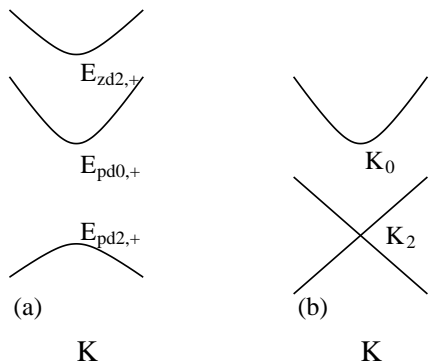}
\caption{Schematic band structure close to the K point
for the valence and conduction bands: (a) including S $p_x$, $p_y$
orbitals; (b) omitting S $p_x$, $p_y$
orbitals.
}
\label{f-dirac}
\end{figure}

In single-layer case, focusing only on the band edges determined
by the states (A) and (B) at the K point, we can identify them with the eigenstates
$E_{pd_0,+}(K)$, $E_{pd_2,+}(K)$, respectively, with a dominant Mo $4d$
character and a marginal S $p_{x/y}$ component, as we show below.
It is thus tempting to define a reduced 3-band tight-binding model,
keeping only the Mo $4d_{3z^2-r^2}$, $4d_{x^2-y^2}$, $4d_{xy}$  orbitals 
with dominant character and disregarding the S $p_x$, $p_y$ orbitals,
with a small marginal weight.
A similar phenomenological model was proposed in
Ref. \onlinecite{xiao}.
However, the full microscopic description here exposed permits to
point out the inconsistency of such a model.
This can be shown by looking at Eq. (\ref{hk66}).
The band gap at K in the full tight-binding model including S $p_x$, $p_y$ orbitals
is determined by the upper eigenstate of $\hat{H}_{pd_0}$,
\begin{eqnarray}
E_{pd_0,+}(K)
&=&
\frac{K_0+K_p^{\rm E}}{2}
\nonumber\\
&&
+
\sqrt{\left(\frac{K_0-K_p^{\rm E}}{2}\right)^2+4K_{pd_0}^2},
\end{eqnarray}
and the upper eigenstate of $\hat{H}_{pd_2}$,
\begin{eqnarray}
E_{pd_2,+}(K)
&=&
\frac{K_2+K_p^{\rm E}}{2}
\nonumber\\
&&
+
\sqrt{\left(\frac{K_2-K_p^{\rm E}}{2}\right)^2+8K_{pd_2}^2},
\end{eqnarray}
both with main Mo $4d$ character, while the eigenstate
\begin{eqnarray}
E_{zd_2,+}(K)
&=&
\frac{K_2+K_z^{\rm E}}{2}
\nonumber\\
&&
+
\sqrt{\left(\frac{K_2-K_z^{\rm E}}{2}\right)^2+4K_{zd_2}^2},
\end{eqnarray}
also with dominant Mo $4d$ character, but belonging
to the block $\hat{H}_{zd_2}$, lies at higher energy (see table \ref{t-block}).
The 3-band model retaining only the $d_0$, $d_2$ orbitals
is equivalent to switch off the hybridization terms $K_{pd_0}$, $K_{pd_2}$, $K_{zd_2}$, 
ruled by $V_{pd\sigma}$, $V_{pd\pi}$, so that $E_{pd_0,+}(K)=K_0$,
$E_{pd_2,+}(K)=E_{zd_2,+}(K)=K_2$.
In this context the level $E_{zd_2,+}(K)$ becomes
degenerate with $E_{pd_2,+}(K)$. This degeneracy is not accidental but
it reflects the fact that the elementary excitations of the $d_2$
states, in this simplified model, are described by a Dirac spectrum,
as sketched in Fig. \ref{f-dirac}.
As a consequence, no direct gap can be possibly established in this framework.
It is worth to mention that
a spin-orbit coupling can certainly split the Dirac cone to produce a
direct gap at the K point, but it would not explain in any case the
direct gap observed in the DFT calculations without spin-orbit coupling.

We should also mention that, in the same reduced 3-band model
keeping only the $d_0$ and $d_2$ Mo orbitals,
the secondary maximum (C) of the valence band 
would have a pure $d_0$ orbital character.
As we are going to see in the discussion concerning the multilayer
samples,
this would have important consequences on the construction of a proper
tight-binding model.

A final consideration concerns the orbital character of the valence
band edge, $E_{pd_2,+}(K)$. This state is associated with the third
$2 \times 2$ block of (\ref{hk66}) and it results from the
hybridization of the chiral state
$d_{k,R2}=(d_{k,x^2-y^2}+id_{k,xy})/\sqrt{2}$ of the 
Mo $d$ orbitals with the chiral state
$p_{k,R,S}=(p_{k,x,S}+ip_{k,y,S})/\sqrt{2}$
of the S $p$ orbitals.
The role of the chirality associated with the $d$ orbitals,
in the presence of a finite spin-orbit coupling, has been
discussed in detail in relation with spin/valley selective
probes.\cite{mak12,mak13,cao,sallen,xiao,wu,zeng}
What results from a careful tight-binding description is that such
$d$-orbital chirality is indeed entangled with a corresponding chirality
associated with the S $p$ orbitals.
The possibility of such entanglement, dictated by group theory,
was pointed out in Ref. \onlinecite{OR13}.

A similar feature is found for the conduction band edge, $E_{pd_0,+}(K)$.
So far, this state has been assumed to be mainly characterized by the
$d_{3z^2-r^2}$, and hence without an orbital moment.
However, as we can see, this is true only for the Mo $d$ part, whereas
the S $p$ component does contain a finite chiral moment.
On the other hand, the spin-orbit associated with the S atoms as well
as with other chalcogenides (ex.: Se) is quite small, and taking into
account also the small orbital S weight, the possibility of
a direct probe of such orbital moment is still to be explored.

\section{Bulk system}
\label{s:bulk}

In the previous section we have examined in detail the content of
the orbital character in the main high-symmetry points of the Brillouin zone
of the single-layer MoS$_2$, to provide theoretical constraints on the
construction of a suitable tight-binding model.
Focusing on the low-energy excitations close to the direct gap at the
K point, we have seen that a proper model must take into account
at least the three Mo orbitals $d_{3z^2-r^2}$, $d_{x^2-y^2}$, $d_{xy}$
and the two S orbitals $p_x$, $p_y$.
On the other hand, our wider aim is to introduce a tight-binding
model for the single-layer that would be the basilar ingredient
for a tight-binding model in {\em multilayer} systems,
simply adding the interlayer coupling.

For the sake of simplicity we focus here
on the bulk 2H-MoS$_2$ structure as a representative case
that contains already all the ingredients
of the physics of multilayer compounds.
The band structure for the bulk compound
is shown in Fig. \ref{f-bandsDFTmos2bulk}.
As it is known, the secondary maximum (C) of valence band
at the $\Gamma$ point is shifted
to higher energies in multilayer systems
with respect to the single-layer case, becoming the valence band
maximum.
At the same time also the secondary minimum (D) of the conduction band,
roughly at the Q point, is lowered in energy, becoming the
conduction band minimum.
All these changes result in a transition between a direct gap
material in single-layer compounds to indirect gap systems
in the multilayer case.
Although such intriguing feature has been discussed extensively
and experimentally observed, the underlying mechanism
has not been so far elucidated.
We will show here that such topological transition of the band edges can be
naturally explained within the context of a tight-binding model
as a result of an orbital selective (and hence momentum dependent)
band splitting induced by the interlayer hopping.

The orbital content of the bulk band structure
along the same high-symmetry lines as in the single-layer case
is shown in 
Fig. \ref{f-bandsDFTmos2bulk}.
We will focus first on the K point, where the single-layer system has
a direct gap.
We  note that the direct gap at K is hardly affected.
The interlayer coupling produces just a very tiny splitting of
the valence band edge $E_{pd_2,+}(K)$, while the conduction band edge $E_{pd_0,+}(K)$
at K becomes doubly degenerate.

Things are radically different at the $\Gamma$ point.
The analysis of the orbital weight $d_{3z^2-r^2}$ in Fig. \ref{f-bandsDFTmos2bulk}
shows indeed that there is a sizable splitting of the $E_{zd_0,+}(\Gamma)$ level, of the order of $1$ eV.
A bit more difficult to discern, because of the multi-orbital component,
but still visible, is the splitting of the secondary minimum (D) of the
conduction band in Q. This is clearest detected by looking
in Fig. \ref{f-bandsDFTmos2bulk} at the $d_2$ and $d_0$ characters,
which belong unically to the E block.
One can thus estimate from DFT a splitting
of this level at the Q point of $\sim 1.36$ eV.

We are now going to see that all these features are consistent with a tight-binding construction
where the interlayer hopping acts as an additional parameter with
respect to the single-layer tight-binding model.
From the tight-binding point of view, it is clear that the main
processes to be included are the interlayer hoppings between the
external S planes of each MoS$_2$ block.
This shows once more the importance of including the S $p$ orbital
in a reliable tight-binding model.
Moreover, for geometric reasons, one could expect that
the interlayer hopping between the $p_z$ orbitals, pointing directly
out-of-plane, would be dominant with respect to the interlayer hopping
between $p_x$, $p_y$.
This qualitative argument is supported by the DFT results, which
indeed report a big splitting of the $E_{zd_0,+}(\Gamma)$ level
at the $\Gamma$ point, with a 27 \% of $p_z$ component,
but almost no splitting of the degenerate $E_{pd_2,+}(\Gamma)$
at $\sim 2$ eV, with 68 \% component of $p_{x}$, $p_y$.

We can quantify this situation within the tight-binding description by
including explicitly the interlayer hopping between the 
$p$-orbitals of the S atoms in the outer planes
of each MoS$_2$ layer, with interatomic distance $d=3.49$ \AA\,
(see Fig. \ref{f-structure}).
These processes will be parametrized in terms
of the {\rm interlayer} Slater-Koster ligands $U_{pp\sigma}$, $U_{pp\pi}$.
The Hilbert space is now determined by a 22-fold vector, defined as:
\begin{eqnarray}
\tilde{\Phi}_k^\dagger
&=&
(
\tilde{\phi}_{k,1}^\dagger,
\tilde{\phi}_{k,2}^\dagger
),
\label{tilde2}
\end{eqnarray}
where $\tilde{\phi}_{k,1}^\dagger$ represents the basis (\ref{basistilde})
for the layer 1, and $\tilde{\phi}_{k,2}^\dagger$ the same quantity for
the layer 2.
The corresponding Hamiltonian, in the absence of interlayer hopping,
would read thus:
\begin{eqnarray}
\hat{H}_{\rm bulk}
&=&
\left(
\begin{array}{cc}
\hat{H}_1 & \hat{0} \\
\hat{0} & \hat{H}_2
\end{array}
\right),
\label{hmatrbulk0}
\end{eqnarray}
where $\hat{H}_1$, $\hat{H}_2$ refer to the intralayer Hamiltonian
for the layer 1 and 2, respectively.

Note that the Hamiltonian of layer 2 in the 2H-MoS$_2$ structure
is different with respect to the one of layer 1.
From a direct inspection we can see that the elements
$H_{2,\alpha,\beta}(\xi,\eta)$ of layer 2 are related to the corresponding
elements of layer 1 as:
\begin{eqnarray}
H_{2,\alpha,\beta}(\xi,\eta)
=
P_\alpha P_\beta H_{1,\alpha,\beta}(\xi,-\eta),
\label{trans}
\end{eqnarray}
where $\xi=k_xa/2$, $\eta=\sqrt{3}k_ya/2$, and $P_{\alpha}=1$ if the orbital $\alpha$
has even symmetry for $y \rightarrow -y$,
and $P_{\alpha}=-1$ if it has odd symmetry.
We note that both effects can be re-adsorbed in a different
redefinition of the orbital basis so that
the eigenvalues of $\hat{H}_2$ are of course the same as
the eigenvalues of $\hat{H}_1$.

Taking into account the inter-layer S-S hopping terms,
we can write thus:
\begin{eqnarray}
\hat{H}_{\rm bulk}
&=&
\left(
\begin{array}{cc}
\hat{H}_1 & \hat{H}_\perp \\
\hat{H}_\perp^\dagger & \hat{H}_2
\end{array}
\right),
\label{hmatrbulk}
\end{eqnarray}
where $\hat{H}_\perp$
is here the interlayer hopping Hamiltonian, namely:
\begin{eqnarray}
\hat{H}_\perp
&=&
\left(
\begin{array}{cc}
\hat{I}_{\rm E}\cos\zeta & \hat{I}_{\rm EO} \sin\zeta \\
-\hat{I}_{\rm EO}^{\rm T}\sin\zeta & \hat{I}_{\rm O} \cos\zeta
\end{array}
\right),
\label{hperpbulk}
\end{eqnarray}
where $\zeta=k_zc/2$ and
\begin{eqnarray}
\hat{I}_{\rm E}
&=&
\left(
\begin{array}{cc}
\hat{0}_{3 \times 3} & \hat{0}_{3 \times 3} \\
\hat{0}_{3 \times 3} & \hat{I}
\end{array}
\right),
\end{eqnarray}
\begin{eqnarray}
\hat{I}_{\rm O}
&=&
\left(
\begin{array}{cc}
\hat{0}_{2 \times 2} & \hat{0}_{2 \times 3} \\
\hat{0}_{3 \times 2} & \hat{I}
\end{array}
\right),
\end{eqnarray}
\begin{eqnarray}
\hat{I}_{\rm EO}
&=&
\left(
\begin{array}{cc}
\hat{0}_{3 \times 2} & \hat{0}_{3 \times 3} \\
\hat{0}_{3 \times 2} & i\hat{I}
\end{array}
\right),
\end{eqnarray}
\begin{eqnarray}
\hat{I}
&=&
\left(
\begin{array}{ccc}
I_{x/x} & I_{x/y} & I_{x/z} \\
I_{x/y} & I_{y/y} & I_{y/z} \\
I_{x/z} & I_{y/z} & I_{z/z} 
\end{array}
\right).
\label{Ibulk}
\end{eqnarray}

The analytical expression of the elements $I_{\alpha/\beta}$
as functions of the Slater-Koster interlayer parameters $U_{pp\sigma}$, $U_{pp\pi}$
is provided in Appendix \ref{app_elements}.
Note that, in the presence of interlayer hopping in the bulk
MoS$_2$, we cannot divide anymore,
for generic momentum ${\bf k}$, the $22 \times 22$
Hamiltonian in smaller blocks with even and odd symmetry with respect
to the change $z \rightarrow -z$.
The analysis is however simplified at specific high-symmetry points
of the Brillouin zone.
In particular, for $k_z=0$ ($\zeta=0$), we can easily see from
(\ref{hperpbulk}) that the block $12 \times 12$
($6 \times 6+6 \times 6$) with even symmetry and
the block $10 \times 10$
($5 \times 5+5 \times 5$) with odd symmetry are still decoupled.

Exploiting this feature, we can now give a closer look at the
high-symmetry points.

\subsection{$\Gamma$ point}

In Section \ref{s:1L} we have seen that at the $\Gamma$ point
the Hamiltonian can be decomposed in $2 \times 2$ blocks.
Particularly important here is the block $H_{zd_0}$ whose upper
eigenvalue $E_{zd_0,+}(\Gamma)$, with main orbital character
$d_{3z^2-r^2}$ and a small $p_z$ component,
represents the secondary maximum (C) of the valence band.
A first important property to be stressed in bulk systems
is that, within this (Mo $4d$)+(S $3p$) tight-binding model,
the interlayer coupling at the $\Gamma$ point does not mix
any additional orbital character.
This can be seen by noticing that the interlayer matrix $\hat{I}$
is diagonal at the $\Gamma$ point.
Focusing on the $E_{zd_0}(\Gamma)$ levels, we can write thus
a $4 \times 4$ reduced Hamiltonian (see Appendix \ref{app_blocks}):
\begin{eqnarray}
\hat{H}_{zd_0}
&=&
\left(
\begin{array}{cccc}
\Gamma_0 & \sqrt{2}\Gamma_{zd_0} & 0  & 0\\
\sqrt{2}\Gamma_{zd_0} & \Gamma_{z}^{\rm E} & 0 & \Gamma_{zz} \\
 0 & 0 & \Gamma_0 & \sqrt{2}\Gamma_{zd_0} \\
 0 & \Gamma_{zz} & \sqrt{2}\Gamma_{zd_0} & \Gamma_z^{\rm E}
\end{array}
\right),
\label{Hzd_0bulk}
\end{eqnarray}
where $\Gamma_{zz}$ represents the interlayer hopping mediated
by $U_{pp\sigma}$, $U_{pp\pi}$
between $p_z$
orbitals belonging to the outer S planes on different layers.
Eq. (\ref{Hzd_0bulk}) is important because it shows that the
qualitative idea that each energy level in the bulk system is just
split by the interlayer hopping is well grounded.
In particular, under the reasonable hypothesis that the interlayer
hopping is much smaller than intralayer processes,
denoting $E_{zd_0,+a}(\Gamma)$, $E_{zd_0,+b}(\Gamma)$ the two
eigenvalues with primary $d_0$ components, we get:
\begin{eqnarray}
\Delta E_{zd_0,+}(\Gamma)
&=&
E_{zd_0,+a}(\Gamma)-E_{zd_0,+b}(\Gamma)
\nonumber\\
&\approx&
\Gamma_{zz}
\left[
\frac{\Gamma_0-\Gamma_z^{\rm E}}
{2\sqrt{
\displaystyle \left(\frac{\Gamma_0-\Gamma_z^{\rm E}}{2}\right)^2+2\Gamma_{zd_0}}}
-1
\right]
\nonumber\\
&=&
\Gamma_{zz}
\left[
\frac{\Gamma_0-\Gamma_z^{\rm E}}
{E_{zd_0,+}(\Gamma)-E_{zd_0,-}(\Gamma)
}
-1
\right].
\end{eqnarray}

A similar situation is found for the other $2 \times 2$ blocks
$\hat{H}_{pd_2}(\Gamma)$, $\hat{H}_{pd_1}(\Gamma)$,
and the  $1 \times 1$ block $\hat{H}_{z}(\Gamma)$.
Most important, tracking the DFT levels by means of their orbital
content, we can note that both levels $E_{zd0,+}(\Gamma)$ and
$E_{zd0,-}(\Gamma)$  undergo
a quite large splitting $\approx 1.2$ eV,
and the level $E_z(\Gamma)$ a splitting $\approx 2.6$ eV,
whereas the levels $\hat{H}_{pd_2}(\Gamma)$, $\hat{H}_{pd_1}(\Gamma)$
are almost unsplit. This observation strongly suggest that, as
expected, the interlayer hopping between $p_x$, $p_y$ orbitals is much less
effective  than the interlayer hopping between $p_z$.

Similar conclusion can be drawn from the investigation of the energy levels
at the K point, although the analysis is a bit more involved.

\subsection{K point}

The properties of the bulk system at the K point
are dictated by the structure of the interlayer matrix $\hat{I}$
which, in the basis defined in Eq. (\ref{tilde2}),
at the K point reads:
\begin{eqnarray}
\hat{I}_{66}(K)
&=&
\left(
\begin{array}{ccc}
K_{pp} & iK_{pp} & iK_{pz} \\
iK_{pp} & -K_{pp} & K_{pz} \\
iK_{pz} & K_{pz} & 0 
\end{array}
\right).
\end{eqnarray}

As discussed in detail in Appendix \ref{app_blocks},
the electronic structure is made more transparent by using an
appropriate chiral basis, which is a direct generalization of the
one for the single-layer.
We can thus write
the even and odd parts of the resulting Hamiltonian in the form:
\begin{eqnarray}
\hat{H}_{\rm E}(K)
&=&
\left(
\begin{array}{ccc}
\hat{H}_{pzd_{02}}(K) & 0 & 0 \\
0 & \hat{H}_{pzd_{02}}(K) & 0 \\
0 & 0 & \hat{H}_{pd_2,\rm E}(K)
\end{array}
\right),
\label{HEbulk}
\\
\hat{H}_{\rm O}(K)
&=&
\left(
\begin{array}{ccc}
\hat{H}_{pzd_1}(K) & 0 & 0 \\
0 & \hat{H}_{pzd_1}(K) & 0 \\
0 & 0 & \hat{H}_{pd_1,\rm O}(K)
\end{array}
\right),
\end{eqnarray}
where
\begin{widetext}
\begin{eqnarray}
\hat{H}_{pzd_{02}}(K)
&=&
\left(
\begin{array}{cccc}
K_0 & -2iK_{pd_0} & 0 & 0  \\
2iK_{pd_0} & K_p^{\rm E} & 0 &  i\sqrt{2}K_{pz}\\
0  & 0 & K_2 & 2K_{zd_2}  \\
0 & -i\sqrt{2}K_{pz} & 2K_{zd_2} & K_z^{\rm E}
\end{array}
\right),
\\
\hat{H}_{pd_2,\rm E}(K)
&=&
\left(
\begin{array}{cccc}
K_2 & i\sqrt{8}K_{pd_2}  & 0 & 0 \\
-i\sqrt{8}K_{pd_2} & K_p^{\rm E} & 0 & 2K_{pp} \\
0 & 0 & K_2 & i\sqrt{8}K_{pd_2}  \\
0 & 2K_{pp} & -i\sqrt{8}K_{pd_2} & K_p^{\rm E} 
\end{array}
\right),
\label{epd_2bulk}
\\
\hat{H}_{pzd_1}(K)
&=&
\left(
\begin{array}{ccc}
K_1 & -2iK_{zd_1} & 0  \\
2iK_{zd_1} & K_z^{\rm O} & 0  \\
0  & 0 & K_p^{\rm O} 
\end{array}
\right),
\\
\hat{H}_{pd_1,\rm O}(K)
&=&
\left(
\begin{array}{cccc}
K_1 & \sqrt{8}K_{pd_1} & 0 & 0  \\
\sqrt{8}K_{pd_1}  & K_p^{\rm O} & 0 & 2K_{pp} \\
0 & 0 & K_1 & \sqrt{8}K_{pd_1} \\
0 & 2K_{pp} & \sqrt{8}K_{pd_1}  & K_p^{\rm O}
\end{array}
\right).
\label{hpd1O}
\end{eqnarray}
\end{widetext}

We can notice that Eq. (\ref{epd_2bulk}) has the same structure
as (\ref{Hzd_0bulk}), with two $2 \times 2$ degenerate sub-blocks
hybridized by a non-diagonal element ($K_{pp}$ in this case).
This results in a splitting of the single-layer levels
$E_{pd_2,+}(K) \rightarrow E_{pd_2,+a}(K), E_{pd_2,+b}(K)$,
$E_{pd_2,-}(K) \rightarrow E_{pd_2,-a}(K), E_{pd_2,-b}(K)$.
The two levels $E_{pd_2,+a}(K)$, $E_{pd_2,+b}(K)$, by looking at their
orbital character, can be identified in DFT results in the small
splitting of the (B) $E_{pd_2,+}(K)$ level, confirming once more
the smallness of the interlayer $p_{x/y}$-$p_{x/y}$ hopping.

Less straightforward is the case of the $4\times 4$ block
$\hat{H}_{pzd_{02}}(K)$
where the hybridization term $\sqrt{2}K_{pz}$ mixes
two different $2\times 2$  sub-blocks, $\hat{H}_{pd_0}$
and $\hat{H}_{zd_2}$. In this case, a mixing of the orbital character
will result.
We note, however, that the block  $\hat{H}_{pzd_{02}}(K)$
appears twice in (\ref{HEbulk}), so that each energy level
will result double-degenerate, in particular the minimum (A)
of the conduction band at K.
Note, however, that the negligible shift of such energy level
in the DFT calculations with
respect to the single-layer case is an indication that also
the interlayer hopping element $K_{pz}$, between $p_z$ on one layer
and $p_x$, $p_y$ on the other one, is negligible.

\subsection{Q point}

An analytical insight on the electronic structure at the Q point was
not available in single-layer systems and it would be thus even
more complicate in the bulk case.
A few important considerations, concerning the minimum (D),
can be however drawn from the DFT results.
In particular, we note that in the single-layer case this energy
level had a non-vanishing $p_z$ component.
As we have seen above, the interlayer hopping between $p_z$ orbitals
appears to be dominant with respect to the interlayer hopping between
$p_{x/y}$ and $p_{x/y}$ and with respect to the mixed interlayer
hopping $p_z$-$p_{x/y}$. We can thus expect a finite sizable splitting
of the (D) level, containing a finite $p_z$ component, with respect
to the negligible energy shift of $E_{pd_0,+}$ (A), which depends
on the mixed interlayer process $K_{pz}$.

\section{Momentum/orbital selective splitting
and comparison with DFT data}
\label{s:disc}

In the previous section we have elucidated, using a tight-binding
model, the orbital  character of the band structure of MoS$_2$ on the main
high-symmetry points of the Brillouin zone.
We have shown how a reliable minimal model for the single-layer case
needs to take into account at least the $p_x$, $p_y$ orbitals of the S
atoms in addition to the $4d$ orbitals of Mo.
A careful inspection of the electronic structure  shows also that
the band edges at the K point defining the direct band gap in the single-layer
case are characterized not only by a chiral order of the $d$ Mo
orbitals, as experimentally observed, but also by an entangled
chiral order of the minor component of the $p_{x/y}$ S orbitals.

An important role is also played by the $p_z$ orbitals of the S atoms.
In single-layer systems, the $p_z$ orbital character is particularly
relevant in the (C) state, characterizing a secondary maximum
in the valence band at the $\Gamma$ point, and in the (D) state,
which instead provides a secondary minimum
in the conduction band at the Q point.

The $p_z$ component becomes crucial in multilayer compounds
where a comparison with DFT results shows that the interlayer coupling
is mainly driven by the $p_z$-$p_z$ hopping whereas
$p_{x/y}$-$p_{x/y},$ $p_z$-$p_{x/y}$ are negligible.
This results in an orbital-selective and momentum-dependent
interlayer splitting of the energy levels, being larger for the
(C) and (D) states and negligible for (A) and (B).
This splitting is thus the fundamental mechanism responsible
for the transition from a direct (A)-(B) gap in single-layer compounds
to an indirect (C)-(D) gap in multilayer systems.
Controlling these processes is therefore of the highest importance
for electronic applications.
Note that such direct/indirect gap switch is discussed here in terms
of the number of layers. On the other hand, the microscopical
identification of such mechanism, which is essentially driven
by the interlayer coupling, permits to understand on the physical
ground the high sensitivity to pressure/strain effects,
as well as to the temperature, via the lattice expansion.

Finally,
in order to show at a quantitative level how the orbital content
determines the evolution of the electronic structure
from single-layer to multilayer compounds,
we have performed a fitting procedure to determine
the tight-binding parameters that best reproduce the DFT bands 
within the model defined here.
The task was divided in two steps: $i$) we first focus on the
single-layer case to determine the relevant Slater-Koster intra-layer
parameters in this case; $ii$) afterwards, keeping fixed the
intralayer parameters, we determine the interlayer parameters.
To this purpose we employ a simplex method\cite{numrecipes}  
to minimize a weighted mean square error $f_{\rm wMSE}$
between the TB and DFT band energies, defined as
\begin{equation}
f_{\rm wMSE} = 
\sum_{{\bf k},i} w_i({\bf k})
\left[ \epsilon_i^{\rm TB}({\bf k}) - \epsilon_i^{\rm DFT}({\bf k}) \right]^2,
\end{equation} 
where $\epsilon_i^{\rm DFT}({\bf k})$ is the dispersion on the
$i$-th band of the 11 band block under consideration,
$\epsilon_i^{\rm TB}({\bf k})$ the corresponding tight-binding
description,
and $w_i({\bf k})$ a band/momentum resolved weight which can
be used to improve fitting over particular ${\bf k}$-regions
or over selected bands.
In spite of many efforts, we could not find a reliable
fit for the whole electronic structure including the seven valence bands
and the four lowest conduction bands.\cite{noteTB}
As our analysis and our main objective concerns the description of
the valence and conduction bands that define the band gap
of these systems, we focus on finding a set of parameters
that describe properly these bands. 
Since both the lowest conduction and highest valence band 
belong to the electronic states with
even  $z \rightarrow -z$ symmetry, the fit was performed in the
$6 \times 6$ orbital space defined by this symmetry.
In addition, due to the degeneracy at the $\Gamma$ point
and to the band crossing along the $\Gamma$-M direction,
the two conduction bands with even symmetry for $z \rightarrow -z$
were considered in the fit.
Additionally, we give a larger weight to the (A)-(D) band edges in order 
to obtain a better description of the most important features of the band structure.

%
\begin{figure}[t]
\begin{center}
\includegraphics[scale=0.4,clip=]{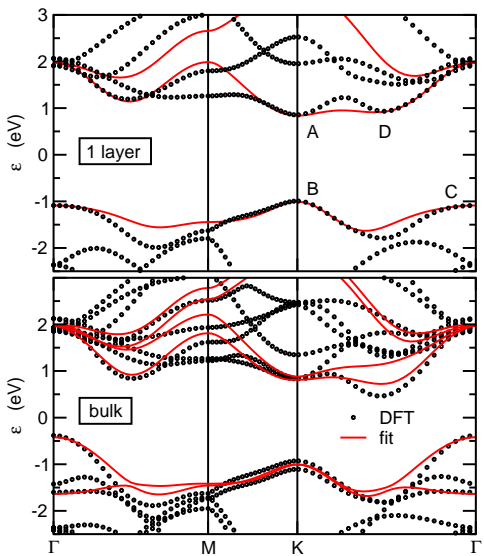}
\end{center}
\caption{Comparison between the DFT band structure
(black dots) and the best fit tight-binding model (red solid lines)
for single-layer (top panel) and bulk MoS$_2$ (bottom panel).
}
\label{f-fit}
\end{figure}

Our best fit for the single-layer case is shown in the top panel
of Fig. \ref{f-fit} (where only the TB bands
with even symmetry $z \rightarrow -z$ are shown),
compared with the DFT bands, and the corresponding
tight-binding parameters are listed in Table \ref{t-fit}.
\begin{table}[b]
\begin{tabular}{lclcr}
\hline
\hline
   Crystal Fields & \hspace{0.5truecm} &$\Delta_0$ & \hspace{0.5truecm} &  -1.016 \\
                          & \hspace{0.5truecm} &$\Delta_1$ & \hspace{0.5truecm} &  -- \\
                          & \hspace{0.5truecm} &$\Delta_2$ & \hspace{0.5truecm} & -2.529 \\
                          & \hspace{0.5truecm} &$\Delta_p$ & \hspace{0.5truecm} &  -0.780 \\
                          & \hspace{0.5truecm} &$\Delta_z$ & \hspace{0.5truecm} &  -7.740 \\
\\                
Intralayer Mo-S & \hspace{0.5truecm} &$V_{pd\sigma}$ & \hspace{0.5truecm} &  -2.619 \\
                          & \hspace{0.5truecm} &$V_{pd\pi}$ & \hspace{0.5truecm} &  -1.396 \\
\\                      
Intralayer Mo-Mo& \hspace{0.5truecm} &$V_{dd\sigma}$ & \hspace{0.5truecm} &  -0.933 \\
                          & \hspace{0.5truecm} &$V_{dd\pi}$ & \hspace{0.5truecm} &  -0.478 \\
                          & \hspace{0.5truecm} &$V_{dd\delta}$ & \hspace{0.5truecm} &  -0.442 \\
\\                        
Intralayer S-S & \hspace{0.5truecm} &$V_{pp\sigma}$ & \hspace{0.5truecm} &  0.696 \\
                          & \hspace{0.5truecm} &$V_{pp\pi}$ & \hspace{0.5truecm} &  0.278 \\
\\                         
Interlayer S-S & \hspace{0.5truecm} &$U_{pp\sigma}$ & \hspace{0.5truecm} &  -0.774 \\
                          & \hspace{0.5truecm} &$U_{pp\pi}$ & \hspace{0.5truecm} &  0.123 \\
\hline
\hline
\end{tabular}
\caption{Tight-binding parameters for single-layer MoS$_2$
($\Delta_\alpha$, $V_\alpha$)
as obtained by fitting the low energy conduction and valence bands.
Also shown are the inter-layer hopping parameters $U_\alpha$
relevant for bulk MoS$_2$. All hopping terms $V_\alpha$, $U_\alpha$
and crystal fields $\Delta_\alpha$ are in units of eV.}
\label{t-fit}
\end{table}
Note that, due to the restriction of our fitting procedure to only some 
bands belonging to the
block with even symmetry, the atomic crystal field $\Delta_1$ for the
Mo orbitals $d_{xz}$, $d_{yz}$ (not involved in the fitting procedure)
results undetermined. The fit reported in Fig. \ref{f-fit} agrees in a qualitative way
with the DFT results, showing, in particular, a direct gap
at the K point [(A) and (B) band edges]
and secondary band edges for the valence and conduction bands
lying at the $\Gamma$ (C) and the Q point (D), respectively.

Turning now to the bulk system,
the further step of determining the interlayer hopping parameters
$U_{pp\sigma}$, $U_{pp\pi}$, is facilitated by
the strong indication, from the DFT analysis, of a dominant role
of the interlayer hopping between the $p_z$ orbitals and a negligible
role of the interlayer hopping between the $p_{x/y}$ orbitals.
Focusing on the $\Gamma$ point, these two different hopping processes
are parametrized in terms of the corresponding interlayer
parameters $\Gamma_{zz}$ and $\Gamma_{pp}$,
as discussed in Appendix \ref{app_blocks}.
We can thus approximate
$\Gamma_{pp}=0$, providing a constraint between 
$U_{pp\sigma}$ and $U_{pp\pi}$, and leaving thus only
one effective independent fitting parameter: $\Gamma_{zz}$.
We determine it, and hence $U_{pp\sigma}$ and $U_{pp\pi}$,
by fixing the effective splitting of the $E_{zd_0,+}(\Gamma)$ level
as in the DFT data.
The values of $U_{pp\sigma}$ and $U_{pp\pi}$ found in this way are
also reported in Table \ref{t-fit}, and the resulting band structure
in the lower panel of Fig. \ref{f-fit}, where only the TB bands
with even symmetry $z \rightarrow -z$ are shown.
We stress that the intralayer hoppings
are here taken from the fitting of the single-layer case.
The agreement between the DFT  and the tight-binding
bands is also  qualitatively good in this case.
In particular, we would like to stress the momentum/orbital
selective interlayer splitting of the bands, which is mainly
concentrated at the $\Gamma$ point for the valence band and
at the Q point for the conduction band. This yields to the
crucial transition between a direct gap in single-layer MoS$_2$,
located at the K point, to an indirect gap $\Gamma$-Q in multilayer
systems.

On more quantitative grounds, we can see that, while the interlayer
splitting of the condution level $E_{zd_0,+}(\Gamma)$ is easily
reproduced, the corresponding
splitting of the conduction band at the Q point is somewhat
underestimated in the tight-binding model ($0.42$ eV)
as compared to the DFT data ($1.36$ eV).
This slight discrepancy is probably due to the
underestimation, in the tight-binding model,
of the $p_z$ character of the conduction band at the Q point.
As a matter of fact, the set of TB parameters reported in Table \ref{t-fit}
gives at the Q point of the conduction band,
for the single-layer case, only a 3.8\% of $p_z$ orbital character,
in comparison with the 11\% found by the DFT calculations.
It should be kept in mind, however, that the optimization of the
tight-binding fitting parameters in such a large phase space
(12 free parameters) is a quite complex and not univocal procedure,
and other solutions are possible.
In particular, a simple algebric analysis suggests that
an alternative solution predicting 11\% of $p_z$ character at the Q
point would yield to a corresponding splitting of the order of $1.2$
eV, in quantitative agreement with the DFT data.
A more refined numerical search in the optimization of the tight-binding
parameters, using global minimization techniques, might result
in better comparison with the DFT results and further work along this line
should be of great interest.

\section{Conclusions}
In this paper we have provided an analytic and reliable description
of the electronic properties
of single-layer and multi-layer semiconduting transition-metal
dichalcogenides in terms of a suitable tight-binding model.
We have shown that the band structure of the multilayer 
compounds can be generated from
the tight-binding model for the single-layer system by adding
the few relevant interlayer hopping terms.
The microscopic mechanism for the transition between a direct-gap to an indirect-gap
from single-layer to multi-layer compounds is thus
explained in terms of a momentum/orbital selective
interlayer band splitting, where the orbital $p_z$ component
of the S atoms plays a central role.
The present work provides with a suitable basis
for the inclusion of many-body effects within the context of Quantum Field Theory
and for the analysis of local strain effects related to the modulation
of the Mo-S, Mo-Mo and S-S ligands.

\acknowledgments
F.G. acknowledges financial support from MINECO, Spain,
through grant FIS2011-23713, and the European Union,
through grant 290846. R. R. acknowledges financial support
from the Juan de la Cierva Program (MINECO, Spain).
E.C. acknowledge support from the European FP7 Marie
Curie project PIEF-GA-2009-251904.
J.A.S.-G. and P.O. ackowledge support from Spanish MINECO
(Grants No. FIS2009-12721-C04-01,
No. FIS2012-37549-C05-02, and
No. CSD2007-00050).
J.A.S.-G. was supported by an FPI Fellowship from MINECO.

\begin{appendix}

\section{Tight-binding Hamiltonian elements}
\label{app_elements}

In this Appendix we provide an analyical expression,
in terms of the Slater-Koster parameters, for the several intra-layer and inter-layer matrix elements
that appear in the Hamiltonian of the tight-binding model.
Following Ref. \onlinecite{doran}, it is convenient to introduce few quantities that
account for the moment dispersion within the Brillouin zone,
as functions of the reduced momentum variables $\xi=k_xa/2$,
$\eta=\sqrt{3}k_ya/2$.

We define thus:
\begin{eqnarray}
C_1(\xi,\eta)
&=&
2\cos(\xi)\cos(\eta/3)+\cos(2\eta/3)
\nonumber\\
&&
+
i[2\cos(\xi)\sin(\eta/3)-\sin(2\eta/3)],
\\
C_2(\xi,\eta)
&=&
\cos(\xi)\cos(\eta/3)-\cos(2\eta/3)
\nonumber\\
&&
+
i[\cos(\xi)\sin(\eta/3)+\sin(2\eta/3)],
\\
C_3(\xi,\eta)
&=&
\cos(\xi)\cos(\eta/3)+2\cos(2\eta/3)
\nonumber\\
&&
+
i[\cos(\xi)\sin(\eta/3)-2\sin(2\eta/3)],
\\
d_1(\xi,\eta)
&=&
\sin(\eta/3)-i\cos(\eta/3),
\\
l_1(\xi,\eta)
&=&
\cos(2\xi)+2\cos(\xi)\cos(\eta),
\\
l_2(\xi,\eta)
&=&
\cos(2\xi)-\cos(\xi)\cos(\eta),
\\
l_3(\xi,\eta)
&=&
2\cos(2\xi)+\cos(\xi)\cos(\eta).
\end{eqnarray}

\subsection{Intra-layer hopping terms}

Following Ref. \onlinecite{doran},
the intralayer hopping terms $H_{\alpha,\beta}$ appearing in
Eqs. (\ref{hmatrpp})-(\ref{hmatrdpb}) can be written as:
\begin{eqnarray*}
H_{x/x}(\xi,\eta)
&=&
\Delta_p+E_{15}l_3(\xi,\eta)+3E_{16}\cos(\xi)\cos(\eta),
\\
H_{y/y}(\xi,\eta)
&=&
\Delta_p+E_{16}l_3(\xi,\eta)+3E_{15}\cos(\xi)\cos(\eta),
\\
H_{z/z}(\xi,\eta)
&=&
\Delta_z
+
2E_{16}l_1(\xi,\eta),
\\
&&
\nonumber\\
\\
H_{z^2/z^2}(\xi,\eta)
&=&
\Delta_0
+
2E_9 l_1(\xi,\eta),
\\
H_{x^2/x^2}(\xi,\eta)
&=&
\Delta_2
+
E_{11} l_3(\xi,\eta)
+
3E_{12}\cos(\xi)\cos(\eta),
\\
H_{xy/xy}(\xi,\eta)
&=&
\Delta_2
+
E_{12} l_3(\xi,\eta)
+
3E_{11}\cos(\xi)\cos(\eta),
\\
H_{xz/xz}(\xi,\eta)
&=&
\Delta_1
+
E_{13} l_3(\xi,\eta)
+
3E_{14}\cos(\xi)\cos(\eta),
\\
H_{yz/yz}(\xi,\eta)
&=&
\Delta_1
+
E_{14} l_3(\xi,\eta)
+
3E_{13}\cos(\xi)\cos(\eta),
\\
&&
\nonumber\\
H_{x/y}(\xi,\eta)
&=&
-\sqrt{3}(E_{15}-E_{16})\sin(\xi)\sin(\eta),
\\
H_{z^2/x^2}(\xi,\eta)
&=&
2E_{10}l_2(\xi,\eta),
\\
H_{z^2/xy}(\xi,\eta)
&=&
-2\sqrt{3}E_{10}\sin(\xi)\sin(\eta),
\\
H_{x^2/xy}(\xi,\eta)
&=&
\sqrt{3}(E_{11}-E_{12})\sin(\xi)\sin(\eta),
\\
H_{xz/yz}(\xi,\eta)
&=&
\sqrt{3}(E_{14}-E_{13})\sin(\xi)\sin(\eta),
\\
&&
\nonumber\\
H_{z^2/x}(\xi,\eta)
&=&
-2\sqrt{3}E_1\sin(\xi) d_1(\xi,\eta),
\\
H_{z^2/y}(\xi,\eta)
&=&
2E_1 C_2(\xi,\eta),
\\
H_{z^2/z}(\xi,\eta)
&=&
E_2 C_1(\xi,\eta),
\\
H_{x^2/x}(\xi,\eta)
&=&
-2\sqrt{3}(\frac{1}{3}E_5-E_3)\sin(\xi) d_1(\xi,\eta),
\\
H_{x^2/y}(\xi,\eta)
&=&
-2E_3C_3(\xi,\eta)-2iE_5\cos(\xi) d_1(\xi,\eta),
\\
H_{x^2/z}(\xi,\eta)
&=&
-2E_4 C_2(\xi,\eta),
\\
H_{xy/x}(\xi,\eta)
&=&
-\frac{2}{3}E_5 C_3(\xi,\eta)-6iE_3\cos(\xi) d_1(\xi,\eta),
\\
H_{xy/y}(\xi,\eta)
&=&
H_{x^2/x}(\xi,\eta),
\\
H_{xy/z}(\xi,\eta)
&=&
2\sqrt{3}E_4\sin(\xi) d_1(\xi,\eta),
\\
H_{xz/x}(\xi,\eta)
&=&
\frac{2}{3}E_6 C_3(\xi,\eta)+6iE_7\cos(\xi) d_1(\xi,\eta),
\\
H_{xz/y}(\xi,\eta)
&=&
2\sqrt{3}(\frac{1}{3}E_6-E_7)\sin(\xi) d_1(\xi,\eta),
\\
H_{xz/z}(\xi,\eta)
&=&
-2\sqrt{3}E_8\sin(\xi) d_1(\xi,\eta),
\\
H_{yz/x}(\xi,\eta)
&=&
H_{xz/y}(\xi,\eta),
\\
H_{yz/y}(\xi,\eta)
&=&
2E_7C_3(\xi,\eta)+2iE_6\cos(\xi) d_1(\xi,\eta),
\\
H_{yz/z}(\xi,\eta)
&=&
2E_8 C_2(\xi,\eta),
\end{eqnarray*}
where
\begin{eqnarray}
E_1
&=&
\frac{1}{2}
\left[
-
V_{pd\sigma}
\left(\sin^2\phi-\frac{1}{2}\cos^2\phi\right)
+
\sqrt{3}
V_{pd\pi}
\sin^2\phi
\right]
\nonumber\\
&&\times
\cos\phi,
\\
E_2
&=&
\left[
-
V_{pd\sigma}
\left(\sin^2\phi-\frac{1}{2}\cos^2\phi\right)
-
\sqrt{3}
V_{pd\pi}
\cos^2\phi
\right]
\nonumber\\
&&\times
\sin\phi,
\\
E_3
&=&
\frac{1}{4}
\left[
\frac{\sqrt{3}}{2}
V_{pd\sigma}
\cos^3\phi
+
V_{pd\pi}
\cos\phi\sin^2\phi
\right],
\\
E_4
&=&
\frac{1}{2}
\left[
\frac{\sqrt{3}}{2}
V_{pd\sigma}
\sin\phi
\cos^2\phi
-
V_{pd\pi}
\sin\phi\cos^2\phi
\right],
\\
E_5
&=&
-
\frac{3}{4}
V_{pd\pi}
\cos\phi,
\\
E_6
&=&
-
\frac{3}{4}
V_{pd\pi}
\sin\phi,
\\
E_7
&=&
\frac{1}{4}
\left[
-\sqrt{3}
V_{pd\sigma}
\cos^2\phi
-
V_{pd\pi}
(1-2\cos^2\phi)
\right]
\nonumber\\
&&\times
\sin\phi,
\\
E_8
&=&
\frac{1}{2}
\left[
-\sqrt{3}
V_{pd\sigma}
\sin^2\phi
-
V_{pd\pi}
(1-2\sin^2\phi)
\right]
\nonumber\\
&&\times
\cos\phi,
\\
&&
\nonumber\\
E_9
&=&
\frac{1}{4}V_{dd\sigma}+\frac{3}{4}V_{dd\delta},
\\
E_{10}
&=&
-\frac{\sqrt{3}}{4}
\left[
V_{dd\sigma}-V_{dd\delta}
\right],
\\
E_{11}
&=&
\frac{3}{4}V_{dd\sigma}+\frac{1}{4}V_{dd\delta},
\\
E_{12}
&=&
V_{dd\pi},
\\
E_{13}
&=&
V_{dd\pi},
\\
E_{14}
&=&
V_{dd\delta},
\\
&&
\nonumber\\
E_{15}
&=&
V_{pp\sigma},
\\
E_{16}
&=&
V_{pp\pi}.
\end{eqnarray}

Here the angle $\phi$ characterize the structure of the
unit cell of the compound and it is determined by purely geometric
reasons (see Fig. \ref{f-structure})..
For the ideal trigonal prism structure, neglecting the
marginal deviations from it in real systems, we have 
$\phi=\arccos[\sqrt{4/7}]$, so that
$\cos\phi=\sqrt{4/7}$ and $\sin\phi=\sqrt{3/7}$.

With these expressions,
taking into account also the further changes of basis, the Hamiltonian
at the $\Gamma$ point
can be divided in sub-blocks as:
\begin{eqnarray}
\hat{H}_{\rm E}(\Gamma)
&=&
\left(
\begin{array}{ccc}
\hat{H}_{zd_0}(\Gamma) & 0 &  0 \\
0 & \hat{H}_{pd_2}(\Gamma) &  0 \\
0 & 0 & \hat{H}_{pd_2}(\Gamma)
\end{array}
\right),
\label{hgammaEapp}
\end{eqnarray}
\begin{eqnarray}
\hat{H}_{\rm O}(\Gamma)
&=&
\left(
\begin{array}{ccc}
\hat{H}_{pd_1}(\Gamma) & 0 &  0 \\
0 & \hat{H}_{pd_1}(\Gamma) &  0 \\
0 & 0 & \Gamma_z
\end{array}
\right),
\label{hgammaOapp}
\end{eqnarray}
where
\begin{eqnarray}
\hat{H}_{zd_0}(\Gamma)
&=&
\left(
\begin{array}{cc}
\Gamma_0 & \sqrt{2}\Gamma_{zd_0}  \\
\sqrt{2}\Gamma_{zd_0} & \Gamma_z^{\rm E}
\end{array}
\right),
\end{eqnarray}
\begin{eqnarray}
\hat{H}_{pd_2}(\Gamma)
&=&
\left(
\begin{array}{cc}
\Gamma_2 & \sqrt{2}\Gamma_{pd_2}  \\
\sqrt{2}\Gamma_{pd_2} & \Gamma_p^{\rm E}
\end{array}
\right),
\end{eqnarray}
\begin{eqnarray}
\hat{H}_{pd_1}(\Gamma)
&=&
\left(
\begin{array}{cc}
\Gamma_1 & \sqrt{2}\Gamma_{pd_2}  \\
\sqrt{2}\Gamma_{pd_2} & \Gamma_p^{\rm O}
\end{array}
\right).
\end{eqnarray}

The parameters $\Gamma_\alpha$ can be viewed as ``molecular'' energy
levels, and the quantities $\Gamma_{\alpha,\beta}$ as hybridization parameters.
Their explicit expressions read:
\begin{eqnarray}
\Gamma_0
&=&
H_{z^2/z^2}(\Gamma)
=
\Delta_0
+
6E_9 ,
\\
\Gamma_1
&=&
H_{xz/xz}(\Gamma)
=
H_{yz/yz}(\Gamma)
\nonumber\\
&=&
\Delta_1
+
3[E_{13}+E_{14}],\
\\
\Gamma_2
&=&
H_{xy/xy}(\Gamma)
=
H_{x^2/x^2}(\Gamma)
\nonumber\\
&=&
\Delta_2
+
3[E_{11}+E_{12}],
\\
\Gamma_p^{\rm E}
&=&
\Gamma_p+V_{pp\pi},
\\
\Gamma_p^{\rm O}
&=&
\Gamma_p-V_{pp\pi},
\\
\Gamma_z^{\rm E}
&=&
\Gamma_z-V_{pp\sigma},
\\
\Gamma_z^{\rm O}
&=&
\Gamma_z+V_{pp\sigma},
\\
\Gamma_p
&=&
H_{x/x}(\Gamma)
=
H_{y/y}(\Gamma)
\nonumber\\
&=&
\Delta_p+3[E_{15}+E_{16}],
\\
\Gamma_z
&=&
H_{z/z}(\Gamma)
=
\Delta_z
+
6E_{16},
\\
\Gamma_{zd_0}
&=&
H_{3z^2-r^2/z}(\Gamma)
=
3E_2 ,
\\
\Gamma_{pd_2}
&=&
H_{x^2-y^2/y}(\Gamma)
=
H_{xy/x}(\Gamma)
\nonumber\\
&=&
-2[3E_3+E_5],
\\
\Gamma_{pd_1}
&=&
H_{xz/x}(\Gamma)
=
H_{yz/y}(\Gamma)
\nonumber\\
&=&
2[3E_7+E_6].
\end{eqnarray}

At the K point, in the proper basis described in the main text,
we can write the even and odd blocks of the Hamiltonian as:
\begin{eqnarray}
\hat{H}_{\rm E}(K)
&=&
\left(
\begin{array}{ccc}
\hat{H}_{pd_0}(K) & 0 &  0 \\
0 & \hat{H}_{zd_2}(K) &  0 \\
0 & 0 & \hat{H}_{pd_2}(K) 
\end{array}
\right),
\\
\hat{H}_{\rm O}
&=&
\left(
\begin{array}{ccc}
\hat{H}_{pd_1}(K) & 0 &  0 \\
0 & \hat{H}_{zd_1}(K) &  0 \\
0 & 0 & K_p^{\rm O}
\end{array}
\right),
\label{hkO}
\end{eqnarray}
where
\begin{eqnarray}
\hat{H}_{pd_0}(K)
&=&
\left(
\begin{array}{cc}
K_0 & -2iK_{pd_0}  \\
2iK_{pd_0} & K_p^{\rm E}
\end{array}
\right),
\end{eqnarray}
\begin{eqnarray}
\hat{H}_{zd_2}(K)
&=&
\left(
\begin{array}{cc}
K_2 & 2K_{zd_2}  \\
2K_{zd_2} & K_z^{\rm E}
\end{array}
\right),
\end{eqnarray}
\begin{eqnarray}
\hat{H}_{pd_2}(K)
&=&
\left(
\begin{array}{cc}
K_2 & i\sqrt{8}K_{pd_2}  \\
-i\sqrt{8}K_{pd_2} & K_p^{\rm E}
\end{array}
\right),
\end{eqnarray}
\begin{eqnarray}
\hat{H}_{pd_1}(K)
&=&
\left(
\begin{array}{cc}
K_1 & \sqrt{8}K_{pd_1}  \\
\sqrt{8}K_{pd_1} & K_p^{\rm O}
\end{array}
\right),
\end{eqnarray}
\begin{eqnarray}
\hat{H}_{zd_1}(K)
&=&
\left(
\begin{array}{cc}
K_1 & -2iK_{zd_1}  \\
2iK_{zd_1} & K_z^{\rm O}
\end{array}
\right).
\end{eqnarray}

The parameters $K_\alpha$, $K_{\alpha,\beta}$ read here:
\begin{eqnarray}
K_0
&=&
H_{z^2/z^2}(K)
=
\Delta_0-3E_9,
\\
K_1
&=&
H_{xz/xz}(K)
=
H_{yz/yz}(K)
\nonumber\\
&=&
\Delta_1
-
\frac{3}{2}
[E_{13}+E_{14}],
\\
K_2
&=&
H_{xy/xy}(K)
=
H_{x^2/x^2}(K)
\nonumber\\
&=&
\Delta_2
-
\frac{3}{2}
[E_{11}+E_{12}],
\\
K_p^{\rm E}
&=&
K_p+V_{pp\pi},
\\
K_p^{\rm O}
&=&
K_p-V_{pp\pi},
\\
K_z^{\rm E}
&=&
K_z-V_{pp\sigma},
\\
K_z^{\rm O}
&=&
K_z+V_{pp\sigma},
\\
K_p
&=&
H_{x/x}(K)
=
H_{y/y}(K)
\nonumber\\
&=&
\Delta_p
-\frac{3}{2}
[E_{15}+E_{16}],
\\
K_z
&=&
H_{z/z}(K)
=
\Delta_z
-3E_{16},
\\
K_{pd_0}
&=&
H_{3z^2-r^2/y}(K)
=
iH_{3z^2-r^2/x}(K)
\nonumber\\
&=&
-3E_1,
\\
K_{zd_2}
&=&
H_{x^2-y^2/z}(K)
=
i H_{xy/z}(K)
\nonumber\\
&=&
3E_4 ,
\\
K_{pd_2}
&=&
H_{x^2-y^2/y}(K)
=
-H_{xy/x}(K)
\nonumber\\
&=&
-i H_{x^2-y^2/x}(K)
=
-i H_{xy/y}(K)
\nonumber\\
&=&
\left[E_5-3E_3\right],
\\
K_{pd_1}
&=&
H_{xz/x}(K)
=
-H_{yz/y}(K)
\nonumber\\
&=&
i H_{xz/y}(K)
=
i H_{yz/x}(K)
\nonumber\\
&=&
[E_6 -3E_7],
\\
K_{zd_1}
&=&
H_{yz/z}(K)
=
i H_{xz/z}(K)
\nonumber\\
&=&
-3E_8.
\end{eqnarray}

\subsection{Inter-layer hopping terms}

Inter-layer hopping is ruled by the Slater-Koster parameters
$U_{pp\sigma}$, $U_{pp\pi}$ describing hopping between S-$3p$ orbitals
belonging to different layers.

In terms of the reduced momentum variables $\xi=k_xa/2$,
$\eta=\sqrt{3}k_ya/2$,
we have thus:
\begin{eqnarray}
I_{x/x}(\xi,\eta)
&=&
\frac{1}{2}\left[E_{19}C_3(\xi,-\eta)+i3E_{17}\cos\xi d_1(\xi,-\eta)\right],
\\
I_{y/y}(\xi,\eta)
&=&
\frac{1}{2}\left[E_{17}C_3(\xi,-\eta)+i3E_{19}\cos\xi d_1(\xi,-\eta)\right],
\\
I_{z/z}(\xi,\eta)
&=&
E_{18}C_1(\xi,-\eta),
\\
I_{x/y}(\xi,\eta)
&=&
\frac{\sqrt{3}}{2}\left[E_{17}-E_{19}\right]\sin\xi d_1(\xi,-\eta),
\\
I_{x/z}(\xi,\eta)
&=&
-\sqrt{3}E_{20}\sin\xi d_1(\xi,\eta),
\\
I_{y/z}(\xi,\eta)
&=&
-E_{20}C_2(\xi,-\eta),
\\
I_{z/z}(\xi,\eta)
&=&
E_{18}C_1(\xi,-\eta),
\end{eqnarray}
where
\begin{eqnarray}
E_{17}
&=&
U_{pp\sigma}\cos^2\beta
+
U_{pp\pi}\sin^2\beta
,
\\
E_{18}
&=&
U_{pp\sigma}\sin^2\beta
+
U_{pp\pi}\cos^2\beta,
\\
E_{19}
&=&
U_{pp\pi},
\\
E_{20}
&=&
\left[U_{pp\sigma}-U_{pp\pi}\right]
\cos\beta\sin\beta.
\end{eqnarray}

Here $\beta$ is the angle between the line connecting the two S atoms
with respect to the S planes (see Fig. \ref{f-structure}).
Denoting $w$ the distance between the two S-planes,
we have:
\begin{eqnarray}
\cos\beta
&=&
\frac{a}{\sqrt{a^2+3w^2}},
\\
\sin\beta
&=&
\frac{\sqrt{3}w}{\sqrt{a^2+3w^2}}.
\end{eqnarray}

Using typical values for bulk MoS$_2$, $a=3.16$ \AA,
and $w=2.975$ \AA, we get $\cos\beta=0.523$ and
$\sin\beta=0.852$.

At the high-symmetry points $\Gamma$, K,
we have thus:
\begin{eqnarray}
\Gamma_{pp}
&=&
I_{x/x}(\Gamma)=
I_{y/y}(\Gamma)
\nonumber\\
&=&
\frac{3}{2}\left[E_{19}+E_{17} \right],
\\
\Gamma_{zz}
&=&
I_{z/z}(\Gamma)
\nonumber\\
&=&
3E_{18},
\\
K_{pp}
&=&
I_{x/x}(K)
=
-I_{y/y}(K)
\nonumber\\
&=&
-iI_{x/y}(K)
=
-iI_{y/x}(K)
\nonumber\\
&=&
\frac{3}{4}\left[E_{19}-E_{17} \right],
\\
K_{pz}
&=&
I_{y/z}(K)
=
I_{z/y}(K)
\nonumber\\
&=&
-iI_{x/z}(K)
=
-iI_{z/x}(K)
\nonumber\\
&=&
\frac{3}{2}E_{20}.
\end{eqnarray}

\section{Decomposition of the Hamiltonian in sub-blocks
at high-symmetry points}
\label{app_blocks}

In this Appenddix we summarize the different unitary transformations
that permit to decomposed at special high-symmetry points
the higher rank Hamiltonian matrix in smaller sub-locks.
In all the cases we treat in a separate way the ``even'' and ``odd''
blocks, namely electronix states with even and odd symmetry with
respect to the $z \rightarrow -z$ inversion.

\subsection{Single-layer}

\subsubsection{$\Gamma$ point}

In the Hilbert space defined by the vector basis $\tilde{\phi}_k^\dagger$
in Eq. (\ref{basistilde}), the even and odd blocks of the Hamiltonian
can be written respectively as:
\begin{widetext}
\begin{eqnarray}
\hat{H}_{\rm E}(\Gamma)
&=&
\left(
\begin{array}{cccccc}
\Gamma_0 & 0 &  0 &  0 & 0 & \sqrt{2}\Gamma_{zd_0} \\
0 & \Gamma_2 &  0 &  0 &\sqrt{2}\Gamma_{pd_2} & 0 \\
0 & 0 &  \Gamma_2 & \sqrt{2}\Gamma_{pd_2} & 0 & 0 \\
0 & 0 &  \sqrt{2}\Gamma_{pd_2} &  \Gamma_p^{\rm E} & 0 & 0 \\
0 & \sqrt{2}\Gamma_{pd_2} &  0 &  0 & \Gamma_p^{\rm E}  & 0 \\
\sqrt{2}\Gamma_{zd_0}  & 0 &  0 &  0 &0 & \Gamma_z^{\rm E} 
\end{array}
\right),
\label{appge}
\end{eqnarray}
and
\begin{eqnarray}
\hat{H}_{\rm O}(\Gamma)
&=&
\left(
\begin{array}{ccccc}
\Gamma_1 & 0 & \sqrt{2}\Gamma_{pd_1} & 0 & 0 \\
0 & \Gamma_1 & 0 & \sqrt{2}\Gamma_{pd_1} & 0 \\
\sqrt{2}\Gamma_{pd_1} & 0 & \Gamma_p^{\rm O} & 0 & 0 \\
0 & \sqrt{2}\Gamma_{pd_1} &0 & \Gamma_p^{\rm O} & 0 \\
0 & 0 & 0 & 0 & \Gamma_z^{\rm O} 
\end{array}
\right).
\label{appgo}
\end{eqnarray}
\end{widetext}

The division in sub-blocks is already evident in
Eqs. (\ref{appge})-(\ref{appgo}).
They can be further ordered using the basis
\begin{eqnarray}
\bar{\phi}_k^\dagger
&=&
(
\bar{\phi}_{k,zd_0}^\dagger,
\bar{\phi}_{k,pd_2,y}^\dagger,
\bar{\phi}_{k,pd_2,x}^\dagger,
\bar{\phi}_{k,pd_1,x}^\dagger,
\bar{\phi}_{k,pd_1,y}^\dagger,
\bar{\phi}_{k,z}^\dagger,
),
\end{eqnarray}
where
\begin{eqnarray}
\bar{\phi}_{k,zd_0}^\dagger
&=&
(
d_{k,3z^2-r^2}^\dagger,
p_{k,z,A}^\dagger,
),
\\
\bar{\phi}_{k,pd_2,y}^\dagger
&=&
(
d_{k,x^2-y^2}^\dagger,
p_{k,y,S}^\dagger
),
\\
\bar{\phi}_{k,pd_2,x}^\dagger
&=&
(
d_{k,xy}^\dagger,
p_{k,x,S}^\dagger
),
\\
\bar{\phi}_{k,pd_1,x}^\dagger
&=&
(
d_{k,xz}^\dagger,
p_{k,x,A}^\dagger
),
\\
\bar{\phi}_{k,pd_1,y}^\dagger
&=&
(
d_{k,yz}^\dagger,
p_{k,y,A}^\dagger
),
\\
\bar{\phi}_{k,z}^\dagger
&=&
(
p_{k,z,S}^\dagger
).
\end{eqnarray}

In this basis we get Eqs. (\ref{hgamma66})-(\ref{hgamma55}),
where
\begin{eqnarray}
\hat{H}_{zd_0}(\Gamma)
&=&
\left(
\begin{array}{cc}
\Gamma_0 & \sqrt{2}\Gamma_{zd_0}  \\
\sqrt{2}\Gamma_{zd_0} & \Gamma_z^{\rm E}
\end{array}
\right),
\\
\hat{H}_{pd_2}(\Gamma)
&=&
\left(
\begin{array}{cc}
\Gamma_2 & \sqrt{2}\Gamma_{pd_2}  \\
\sqrt{2}\Gamma_{pd_2} & \Gamma_p^{\rm E}
\end{array}
\right),
\\
\hat{H}_{pd_1}(\Gamma)
&=&
\left(
\begin{array}{cc}
\Gamma_1 & \sqrt{2}\Gamma_{pd_1}  \\
\sqrt{2}\Gamma_{pd_1} & \Gamma_p^{\rm O}
\end{array}
\right).
\end{eqnarray}

\subsubsection{K point}

In the basis defined by the Hilbert vector $\tilde{\phi}_k^\dagger$,
the Hamiltonian at the K point reads, for the even and odd blocks,
respectively:
\begin{widetext}
\begin{eqnarray}
\hat{H}_{\rm E}(K)
&=&
\left(
\begin{array}{cccccc}
K_0 & 0 &  0 &  -i\sqrt{2}K_{pd_0} &\sqrt{2}K_{pd_0} & 0 \\
0 & K_2 &  0 &  i\sqrt{2}K_{pd_2} & \sqrt{2} K_{pd_2} & \sqrt{2}K_{zd_2} \\
0 & 0 &  K_2 &   -\sqrt{2} K_{pd_2} & i\sqrt{2} K_{pd_2}&
-i\sqrt{2}K_{zd_2} \\
i\sqrt{2}K_{pd_0} & -i\sqrt{2} K_{pd_2} &  -\sqrt{2} K_{pd_2} &
K_p^{\rm E} & 0 & 0 \\
\sqrt{2}K_{pd_0} & \sqrt{2} K_{pd_2} & -i\sqrt{2} K_{pd_2} &  0 &
K_p^{\rm E} & 0 \\
0 & \sqrt{2}K_{zd_2} &  i\sqrt{2}K_{zd_2} &  0 & 0 & K_z^{\rm E} \\
\end{array}
\right),
\label{appke}
\end{eqnarray}
\begin{eqnarray}
\hat{H}_{\rm O}(K)
&=&
\left(
\begin{array}{ccccc}
K_1 & 0 & \sqrt{2}K_{pd_1} & -i\sqrt{2}K_{pd_1} & -i\sqrt{2}K_{zd_1} \\
0 & K_1 & -i\sqrt{2}K_{pd_1} & -\sqrt{2}K_{pd_1} & \sqrt{2}K_{zd_1} \\
\sqrt{2}K_{pd_1} & i\sqrt{2}K_{pd_1} &  K_p^{\rm O} & 0 & 0 \\
i\sqrt{2}K_{pd_1} & -\sqrt{2}K_{pd_1} & 0 & K_p^{\rm O} & 0 \\
i\sqrt{2}K_{zd_1} & \sqrt{2}K_{zd_1} & 0 & 0 & K_z^{\rm O} 
\end{array}
\right).
\label{appko}
\end{eqnarray}
\end{widetext}

In order to decopled the Hamiltoniam, it is convenient to
introduce the chiral basis defined by the vector $\bar{\psi}_k^\dagger$
in (\ref{psi}). In this Hilbert space we have thus:
\begin{eqnarray}
\hat{H}_{\rm E}(K)
&=&
\left(
\begin{array}{ccc}
\hat{H}_{pd_0}(K) & 0 &  0 \\
0 & \hat{H}_{zd_2}(K) &  0 \\
0 & 0 & \hat{H}_{pd_2}(K) 
\end{array}
\right),
\\
\hat{H}_{\rm O}
&=&
\left(
\begin{array}{ccc}
\hat{H}_{pd_1}(K) & 0 &  0 \\
0 & \hat{H}_{zd_1}(K) &  0 \\
0 & 0 & K_p^{\rm O} 
\end{array}
\right),
\end{eqnarray}
where
\begin{eqnarray}
\hat{H}_{pd_0}(K)
&=&
\left(
\begin{array}{cc}
K_0 & -i2K_{pd_0}  \\
i2K_{pd_0} & K_p^{\rm E}
\end{array}
\right),
\\
\hat{H}_{zd_2}(K)
&=&
\left(
\begin{array}{cc}
K_2 & 2K_{pd_2}  \\
2K_{pd_2} & K_z^{\rm E}
\end{array}
\right),
\\
\hat{H}_{pd_2}(K)
&=&
\left(
\begin{array}{cc}
K_2 & i\sqrt{8}K_{pd_2}  \\
-i\sqrt{8}K_{pd_2} & K_p^{\rm E}
\end{array}
\right),
\\
\hat{H}_{pd_1}(K)
&=&
\left(
\begin{array}{cc}
K_1 & \sqrt{8}K_{pd_1}  \\
\sqrt{8}K_{pd_1} & K_p^{\rm O}
\end{array}
\right),
\\
\hat{H}_{zd_1}(K)
&=&
\left(
\begin{array}{cc}
K_1 & -i2K_{pd_1}  \\
i2K_{pd_1} & K_z^{\rm O}
\end{array}
\right).
\end{eqnarray}

\subsection{Bulk system}

The general structure of the tight-binding Hamiltonian $\hat{H}_{\rm bulk}$
for the bulk system, using the basis defined in (\ref{tilde2}),
is provided in Eqs. (\ref{hmatrbulk})-(\ref{Ibulk}),
where we also remind the symmetry property (\ref{trans})
that related the matrix elements of $\hat{H}_2$ to $\hat{H}_1$.

As mentioned in the main text,
for $k_z=0$ the band structure can be
still divided in  two independent blocks with even and odd
symmetry with respect to the transformation $z \rightarrow
-z$.\cite{doran}
Further simplicication are encountered at
the high-symmetry points
$\Gamma$ and K.

\subsubsection{$\Gamma$ point}

We first notice that at the $\Gamma$ point the relation
(\ref{trans}) does not play any role,
i.e. $\hat{H}_2(\Gamma)=\hat{H}_1(\Gamma)$,
where $\hat{H}_1(\Gamma)$ is defined by
Eqs. (\ref{hdecoupled})-(\ref{hgamma55}) in the main text.

The Hamiltonian is thus completely determined  by
the interlayer hopping matrix $\hat{I}$ that at the $\Gamma$ point
reads:
\begin{eqnarray}
\hat{I}(\Gamma)
&=&
\left(
\begin{array}{ccc}
\Gamma_{pp} & 0 & 0 \\
0 & \Gamma_{pp} & 0 \\
0 & 0 & \Gamma_{zz} 
\end{array}
\right).
\end{eqnarray}

A convenient basis to decoupled the Hamiltonian in smaller subblocks
is thus:
\begin{eqnarray}
\bar{\Phi}_k^\dagger
&=&
(
\bar{\Phi}_{k,zd_0}^\dagger,
\bar{\Phi}_{k,pd_2,y}^\dagger,
\bar{\Phi}_{k,pd_2,x}^\dagger,
\bar{\Phi}_{k,pd_1,x}^\dagger,
\bar{\Phi}_{k,pd_1,y}^\dagger,
\bar{\Phi}_{k,z}^\dagger,
),
\end{eqnarray}
where
\begin{eqnarray}
\bar{\Phi}_{k,zd_0}^\dagger
&=&
(
d_{k,3z^2-r^2,1}^\dagger,
p_{k,z,A,1}^\dagger,
d_{k,3z^2-r^2,2}^\dagger,
p_{k,z,A,2}^\dagger
),
\\
\bar{\Phi}_{k,pd_2,y}^\dagger
&=&
(
d_{k,x^2-y^2,1}^\dagger,
p_{k,y,S,1}^\dagger,
d_{k,x^2-y^2,2}^\dagger,
p_{k,y,S,2}^\dagger
),
\\
\bar{\Phi}_{k,pd_2,x}^\dagger
&=&
(
d_{k,xy,1}^\dagger,
p_{k,x,S,1}^\dagger,
d_{k,xy,2}^\dagger,
p_{k,x,S,2}^\dagger
),
\\
\bar{\Phi}_{k,pd_1,x}^\dagger
&=&
(
d_{k,xz,1}^\dagger,
p_{k,x,A,1}^\dagger,
d_{k,xz,2}^\dagger,
p_{k,x,A,2}^\dagger
),
\\
\bar{\Phi}_{k,pd_1,y}^\dagger
&=&
(
d_{k,yz,1}^\dagger,
p_{k,y,A,1}^\dagger,
d_{k,yz,2}^\dagger,
p_{k,y,A,2}^\dagger
),
\\
\bar{\Phi}_{k,z}^\dagger
&=&
(
p_{k,z,S,1}^\dagger,
p_{k,z,S,1}^\dagger,
).
\end{eqnarray}

The resulting total Hamiltonian can be written as:
\begin{eqnarray}
\hat{H}_{\rm bulk}(\Gamma)
&=&
\left(
\begin{array}{cc}
\hat{H}_{\rm E, bulk}(\Gamma) & 0  \\
0 & \hat{H}_{\rm O, bulk}(\Gamma)
\end{array}
\right),
\end{eqnarray}
where
\begin{eqnarray}
\hat{H}_{\rm E, bulk}(\Gamma)
&=&
\left(
\begin{array}{ccc}
\hat{H}_{zd_0,\rm bulk}(\Gamma) & 0 &  0 \\
0 & \hat{H}_{pd_2,\rm bulk}(\Gamma) &  0 \\
0 & 0 & \hat{H}_{pd_2, \rm bulk}(\Gamma)
\end{array}
\right),
\\
\hat{H}_{\rm O, bulk}(\Gamma)
&=&
\left(
\begin{array}{ccc}
\hat{H}_{pd_1,\rm bulk}(\Gamma) & 0 &  0 \\
0 & \hat{H}_{pd_1,\rm bulk}(\Gamma) &  0 \\
0 & 0 & \hat{H}_{z, \rm bulk}(\Gamma)
\end{array}
\right),
\end{eqnarray}
and where
\begin{eqnarray}
\hat{H}_{zd_0,\rm bulk}
&=&
\left(
\begin{array}{cccc}
\Gamma_0 & \sqrt{2}\Gamma_{zd_0} & 0  & 0\\
\sqrt{2}\Gamma_{zd_0} & \Gamma_{z}^{\rm E} & 0 & \Gamma_{zz} \\
 0 & 0 & \Gamma_0 & \sqrt{2}\Gamma_{zd_0} \\
 0 & \Gamma_{zz} & \sqrt{2}\Gamma_{zd_0} & \Gamma_z^{\rm E}
\end{array}
\right),
\\
\hat{H}_{pd_2,\rm bulk}
&=&
\left(
\begin{array}{cccc}
\Gamma_2 & \sqrt{2}\Gamma_{pd_2} & 0  & 0\\
\sqrt{2}\Gamma_{pd_2} & \Gamma_p^{\rm E} & 0 & \Gamma_{pp} \\
 0 & 0 & \Gamma_2 & \sqrt{2}\Gamma_{pd_2} \\
 0 & \Gamma_{pp} & \sqrt{2}\Gamma_{pd_2} & \Gamma_p^{\rm E}
\end{array}
\right),
\\
\hat{H}_{pd_1,\rm bulk}
&=&
\left(
\begin{array}{cccc}
\Gamma_1 & \sqrt{2}\Gamma_{pd_1} & 0  & 0\\
\sqrt{2}\Gamma_{pd_1} & \Gamma_p^{\rm O} & 0 & \Gamma_{pp} \\
 0 & 0 & \Gamma_1 & \sqrt{2}\Gamma_{pd_1} \\
 0 & \Gamma_{pp} & \sqrt{2}\Gamma_{pd_1} & \Gamma_p^{\rm O}
\end{array}
\right),
\\
\hat{H}_{z,\rm bulk}
&=&
\left(
\begin{array}{cc}
\Gamma_z^{\rm O} & \Gamma_{zz}\\
\Gamma_{zz} & \Gamma_z^{\rm O}
\end{array}
\right).
\end{eqnarray}

\subsubsection{K point}

The treatment of the bulk Hamiltonian at the K point,
in order to get a matrix clearly divided in blocks,
is a bit less straighforward than at the $\Gamma$ point.

We first notice that the interlayer matrix,
in the basis $\tilde{\Phi}_k^\dagger$, reads:
\begin{eqnarray}
\hat{I}(K)
&=&
\left(
\begin{array}{ccc}
K_{pp} & iK_{pp} & iK_{pz} \\
iK_{pp} & -K_{pp} & K_{pz} \\
iK_{pz} & K_{pz} & 0 
\end{array}
\right).
\end{eqnarray}

We then ridefine the orbitals
$d_{k,yz,2}^\dagger \rightarrow \bar{d}_{k,yz,2}^\dagger=- d_{k,yz,2}^\dagger$,
$p_{k,y,\alpha,2}^\dagger \rightarrow \bar{p}_{k,y,\alpha,2}^\dagger=- p_{k,y,\alpha,2}^\dagger$
($\alpha=$A,S), in order to get, according with (\ref{trans}),
$\hat{H}_2(\Gamma)=\hat{H}_1(\Gamma)$.

Following what done for the single layer, we can also
introduce here a chiral basis.
After a further rearrangement of the vector elements,
we define thus the convenient Hilbert space as:
\begin{widetext}
\begin{eqnarray}
\bar{\Psi}_{k}^\dagger
&=&
(
\bar{\Psi}_{k,pzd_02,L}^\dagger,
\bar{\Psi}_{k,pzd_02,R}^\dagger,
\bar{\Psi}_{k,pd_2,\rm E}^\dagger,
\bar{\Psi}_{k,pzd_1,R}^\dagger,
\bar{\Psi}_{k,pzd_1,L}^\dagger,
\bar{\Psi}_{k,pd_1,\rm O}^\dagger
),
\label{psibarra}
\end{eqnarray}
\end{widetext}
where
\begin{eqnarray}
\bar{\Psi}_{k,pzd_02,L}^\dagger
&=&
(
d_{k,3z^2-r^2,1}^\dagger,
p_{k,L,S,1}^\dagger,
d_{k,R,2}^\dagger,
p_{k,z,A,2}^\dagger,
),
\\
\bar{\Psi}_{k,pzd_02,R}^\dagger
&=&
(
d_{k,3z^2-r^2,2}^\dagger,
p_{k,R,S,2}^\dagger,
d_{k,L,1}^\dagger,
p_{k,z,A,1}^\dagger,
),
\\
\bar{\Psi}_{k,pd_2,\rm E}^\dagger
&=&
(
d_{k,R,1}^\dagger,
p_{k,R,S,1}^\dagger,
d_{k,L,1}^\dagger,
p_{k,L,S,1}^\dagger,
),
\\
\bar{\Psi}_{k,pzd_1,R}^\dagger
&=&
(
d_{k,R,1}^\dagger,
p_{k,z,S,1}^\dagger,
p_{k,R,A,2}^\dagger
),
\\
\bar{\Psi}_{k,pzd_1,R}^\dagger
&=&
(
d_{k,L,2}^\dagger,
p_{k,z,S,2}^\dagger,
p_{k,L,A,1}^\dagger
),
\\
\bar{\Psi}_{k,pd_1,\rm O}^\dagger
&=&
(
d_{k,L,1}^\dagger,
p_{k,R,A,1}^\dagger,
d_{k,R,2}^\dagger,
p_{k,L,A,2}^\dagger,
).
\end{eqnarray}

In this basis, the Hamiltonian can be once more written as:
\begin{eqnarray}
\hat{H}_{\rm bulk}(K)
&=&
\left(
\begin{array}{cc}
\hat{H}_{\rm E}(K) & 0  \\
0 & \hat{H}_{\rm O}(K)
\end{array}
\right),
\end{eqnarray}
where $\hat{H}_{\rm E}(K)$, $\hat{H}_{\rm O}(K)$
are defined in Eqs. (\ref{HEbulk})-(\ref{hpd1O})
of the main text.

\end{appendix}


\begin{thebibliography}{999}

\bibitem{novoselov1}
K.S. Novoselov, A.K. Geim, S.V. Morozov, D. Jiang, Y. Zhang,
S.V. Dubonos, I.V. Gregorieva, and A.A. Firsov, 
Science {\bf 306}, 666 (2004).

\bibitem{novoselov2}
K.S. Novoselov, D. Jiang, F. Schedin, T.J. Booth, V.V. Khotkivich,
S.V. Morozov, and A.K. Geim,
Proc. Nat. Ac. Sc. {\bf 102}, 10451 (2005).

\bibitem{ZK05}
  Zhang, Yuanbo and Tan, Yan-Wen and Stormer, Horst L and Kim, Philip
  Nature {\bf 438}, 201 (2005).

\bibitem{britnell}
L. Britnell, R.V. Gorbachev, R. Jalil, B.D. Belle, F. Schedin, A. Mishchenko,
T. Georgiou, M.I. Katsnelson, L. Eaves, S.V. Morozov, N.M.R. Peres, J. Leist,
A.K. Geim, K. S. Novoselov, and L. A. Ponomarenko,
Science {\bf 335}, 947 (2012).

\bibitem{li}
T. Li and G. Galli,
J. Phys. Chem. {\bf 111}, 16192 (2007)

\bibitem{lebegue}
S. Leb\`egue and O. Eriksson,
Phys. Rev. B {\bf 79}, 115409 (2009).

\bibitem{mak}
K.F. Mak, C. Lee, J. Hone, J. Shan, and T.F. Heinz,
Phys. Rev. Lett. {\bf 105}, 136805 (2010).

\bibitem{splendiani}
A. Splendiani, L. Sun, Y. Zhang, T. Li, J. Kim,
C.-Y. Chim, G. Galli, and F. Wang,
Nano Lett. {\bf 10}, 1271 (2010).

\bibitem{feng}
J. Feng, X. Qian, C.-W. Huang, and J. Li,
Nature Photon. {\bf 6}, 866 (2012).

\bibitem{lu}
P. Lu, X. Wu, W. Guo, and X.C. Zeng,
Phys. Chem. Chem. Phys. {\bf 14}, 13035 (2012).

\bibitem{pan}
H. Pan and Y.-W. Zhang,
J. Phys. Chem. C {\bf 116}, 11752 (2012).

\bibitem{peelaers}
H. Peelaers and C.G. van de Walle,
Phys. Rev. B {\bf 86}, 241401 (2012).

\bibitem{yun}
W.S. Yun, S.W. Han, S.C. Hong, I.G. Kim, and J.D. Lee,
Phys. Rev. B {\bf 85}, 033305 (2012).

\bibitem{scalise1}
E. Scalise, M. Houssa, G. Pourtois, V. Afanas'ev, and A. Stesmans,
Nano Res. 5, 43 (2012).

\bibitem{scalise2}
E. Scalise, M. Houssa, G. Pourtois, V. Afanas'ev, and A. Stesmans,
Physica E (in press, 2013).

\bibitem{li3}
Y. Li, Y.-l. Li, C.M. Araujo, W. Luo, and R. Ahuja,
arXiv:1211.4052 (2012).

\bibitem{ghorbani}
M. Ghorbani-Asl, S. Borini, A. Kuc, and T. Heine,
arXiv:1301.3469 (2013).

\bibitem{shi}
H. Shi, H. Pan, Y.-W. Zhang, and B.I. Yakobson,
arXiv:1211.5653 (2012).

\bibitem{hromadova}
L. Hromodov\'a, R. Marto\v{n}\'ak, and E. Tosatti,
arXiv:1301.0781 (2013).

\bibitem{mak12}
K.F. Mak, K. He, J. Shan, and T.F. Heinz,
Nature Nanotech. {\bf 7}, 494 (2012).

\bibitem{mak13}
K.F. Mak, K. He, C. Lee, G.H. Lee, J. Hone, T.F. Heinz, and J. Shan,
Nature Mat. {\bf 12}, 207 (2013).

\bibitem{cao}
T. Cao, J. Feng, J. Shi, Q. Niu, and E. Wang,
Nature Commun. {\bf 3}, 887 (2012).

\bibitem{sallen}
G. Sallen, L. Bouet, X. Marie, G. Wang, C.R. Zhu, W.P.Han, Y. Lu,
P.H. Tan, T. Amand, B.L. Liu, and B. Urbaszek,
Phys. Rev. B {\bf 86}, 081301 (2012).

\bibitem{xiao}
D. Xiao, G.-B. Liu, W. Feng, X. Xu, and W. Yao,
Phys. Rev. Lett. {\bf 108}, 196802 (2012).

\bibitem{wu}
S. Wu, J.S. Ross, G.-B. Liu, G. Aivazian, A. Jones, Z. Fei,
W. Zhu, D. Xiao, W. Yao, D. Cobden, and X. Xu,
Nature Phys. {\bf 9}, 149 (2013).

\bibitem{zeng}
H. Zheng, J. Dai, W. Yao, D. Xiao, and X. Cui,
Nature Nanotechn. {\bf 7}, 490 (2012).

\bibitem{OR13}
H. Ochoa and R. Rold{\'a}n,
arXiv:1303.5860 (2013).


\bibitem{tongay}
S. Tongay, J. Zhou, C. Ataca, K. Lo, T.S. Matthews, J. Li,
J.C. Grossman, and J. Wu,
Nano Lett. {\bf 12}, 5576 (2012).

\bibitem{wallace}
P. R. Wallace, Phys. Rev. {\bf 71}, 622 (1947).

\bibitem{reich2002}
S. Reich, J. Maultzsch, C. Thomsen, and P. Ordej\'on, 
Phys. Rev. B {\bf 66}, 035412 (2002).

\bibitem{pacoreview}
For a review see for instance:
A.H. Castro Neto, F. Guinea, N.M.R. Peres,
K.S. Novoselov and A.K. Geim,
Rev. Mod. Phys. {\bf 81}, 109 (2009).

\bibitem{mccann}
E. McCann and M. Koshino,
arXiv:1205.6953 (2012).

\bibitem{nilsson}
J. Nilsson, A.H. Castro Neto, F. Guinea, and N.M.R. Peres,
Phys. Rev. Lett. {\bf 97}, 266801 (2006).

\bibitem{pp1}
B. Partoens and F.M. Peeters,
Phys. Rev. B {\bf 74}, 075404 (2006).

\bibitem{pp2}
B. Partoens and F.M. Peeters,
Phys. Rev. B {\bf 74}, 075404 (2006).

\bibitem{abp}
A.A. Avetisyan, B. Partoens and F.M. Peeters,
Phys. Rev. B {\bf 79}, 035421 (2009); 
Phys. Rev. B {\bf 80}, 195401 (2009); 
Phys. Rev. B {\bf 81}, 115432 (2010).

\bibitem{koshino1}
M. Koshino and E. McCann,
Phys. Rev. B {\bf 79}, 125443 (2009).

\bibitem{koshino2}
M. Koshino,
Phys. Rev. B {\bf 81}, 125304 (2010).

\bibitem{koshino3}
M. Koshino and E. McCann,
Phys. Rev. B {\bf 87}, 045420 (2013).

\bibitem{zhang}
F. Zhang, B. Sahu, H. Min, and A. H. MacDonald,
Phys. Rev. B {\bf 82}, 035409 (2010).

\bibitem{yuan}
S. Yuan, R. Rold\'an, and M.I. Katsnelson,
Phys. Rev. B {\bf 84}, 125455 (2011).

\bibitem{yan}
J.-A. Yan, W.Y. Ruan, and M.Y. Chou,
Phys. Rev. B {\bf 83}, 245418 (2011).

\bibitem{olsen}
R. Olsen, R. van Gelderen, and C. Morais Smith,
Phys. Rev. B {\bf 87}, 115414 (2013).

\bibitem{mak4}
K.F. Mak, M.Y. Sfeir, J.A. Misewich, and T.F. Heinz,
Proc. Nat. Ac. Sc. {\bf 107}, 14999 (2010).

\bibitem{lui}
C.H. Lui, Z.Q. Li, K.F. Mak, E. Cappelluti, and T.F. Heinz,
Nature Phys. {\bf 7}, 944 (2011).

\bibitem{bao}
W. Bao, L. Jing, J. Velasco Jr, Y. Lee, G. Liu, D. Tran, B. Standley, M. Aykol, S.B. Cronin,
D. Smirnov, M. Koshino, E. McCann, M. Bockrath. and C.N. Lau.
Nature Phys. {\bf 7}, 948 (2011).

\bibitem{ubrig}
N. Ubrig, P. Blake, D. van der Marel, and A.B. Kuzmenko,
Europhys. Lett. {\bf 100}, 58003 (2012).


\bibitem{siesta1}
J.M. Soler, E. Artacho, J. Gale, A. Garc\'{\i}a, J. Junquera,
P. Ordej\'on and D. S\'anchez-Portal,
J. Phys.: Condens. Matter {\bf 14}, 2745 (2002).

\bibitem{siesta2}
E. Artacho, E. Anglada, O. Dieguez, J.D. Gale, A. Garc\'{\i}a,
J. Junquera, R.M. Martin, P. Ordej\'on, J. M. Pruneda,
D. S\'anchez-Portal, and J.M. Soler,
J. Phys.: Condens. Matter {\bf 20}, 064208 (2008).

\bibitem{fuhrer}
M. S. Fuhrer, J. Nyg\aa rd, L. Shih, M. Forero, Y.-G. Yoon,
    M. S. C. Mazzoni,    H. J. Choi,    J. Ihm,
    S. G. Louie,    A. Zettl,    and P. L. McEuen, Science {\bf 288}, 494 (2000)
    
\bibitem{wang}
B. Wang, M.-L. Bocquet, S. Marchini, S. Gunther and J. Wintterlin, 
Phisical Chemistry Chemical Physics {\bf 10}, 3530 (2008)

\bibitem{lehtinen}
P.O. Lehtinen, A.S. Foster, Y.C. Ma, A.V. Krasheninnikov and R.M. Nieminen,
Phys. Rev. Lett. {\bf 93} 187202 (2004)

\bibitem{novaes}
F.D. Novaes, R. Rurali and P. Ordej\'on,
ACS Nano  {\bf 4}, 7596 (2010)

\bibitem{ho}
J.H.Ho, Y.H. Lai, Y.H. Chiu, and M.F. Lin,
Nanotech. {\bf 19}, 035712 (2008).

\bibitem{palacios}
J.J. Palacios, J. Fernandez-Rossier, L. Brey,
Phys. Rev. B {\bf 77}, 195428 (2008).

\bibitem{pereira}
A.L.C. Pereira and P.A. Schulz,
Phys. Rev. B {\bf 78}, 125402 (2008).

\bibitem{carpio}
A. Carpio, L.L. Bonilla, F. de Juan, M.A.H. Vozmediano,
New J. Phys. {\bf 10}, 053021 (2008).

\bibitem{lopezsancho}
M.P. L\'opez-Sancho, F. de Juan, M.A.H. Vozmediano,
Phys. Rev. B {\bf 79}, 075413 (2009).

\bibitem{ribeiro}
R.M. Ribeiro, V.M. Pereira, N.M.R. Peres, P.R. Briddon, and A.H. Castro Neto,
New J. Phys. {\bf 11}, 115002 (2009).

\bibitem{yuan1}
S. Yuan, R. Rold\'an, and M.I. Katsnelson,
Phys. Rev. B {\bf 84}, 035439 (2011).

\bibitem{yuan2}
S. Yuan, R. Rold\'an, and M.I. Katsnelson,
Phys. Rev. B {\bf 84}, 125455 (2011).

\bibitem{neek}
M. Neek-Amal, L. Covaci, and F.M. Peeters,
Phys. Rev. B {\bf 86}, 041405 (2012).

\bibitem{yuan3}
S. Yuan, R. Rold\'an, A.-P. Jauho, and M.I. Katsnelson,
Phys. Rev. B {\bf 87}, 085430 (2013).

\bibitem{leconte2010}
N. Leconte, J. Moser, P. Ordej\'on, H. Tao, A. Lherbier, A. Bachtold, F. Alsina, 
C.M. Sotomayor-Torres, J.-C. Charlier and S. Roche, 
ACS Nano {\bf 4}, 4033 (2010)

\bibitem{soriano2011}
D. Soriano, N. Leconte, P. Ordej\'on, J.-C. Charlier, J.J. Palacios and S. Roche,
Phys. Rev. Lett. {\bf 107}, 016602 (2011)

\bibitem{leconte2011}
N. leconte, D. Soriano, S. Roche, P. Ordej\'on, J.-C. Charlier and J.J. Palacios
ACS Nano {\bf 5}, 3987 (2011)

\bibitem{vantuan}
D. Van Tuan, A. Kumar, S. Roche, F. Ortmann, M.F. Thorpe and P. Ordej\'on, 
Phys. Rev. B {\bf R86}, 121408 (2012)


\bibitem{zheng}
H. Zheng, Z.F. Wang, T. Luo, Q.W. Shi, and J. Chen,
Phys. Rev. B {\bf 75}, 165414 (2007).

\bibitem{guinea08}
F. Guinea, M.I. Katsnelson, and M.A.H. Vozmediano,
Phys. Rev. B {\bf 77}, 075442 (2008).

\bibitem{castro1}
E.V. Castro, N.M.R. Peres, J.M.B. Lopes dos Santos,
J. Optoelectron. Adv. Materials {\bf 10}, 1716 (2008).

\bibitem{castro2}
E.V. Castro, M.P. L\'opez-Sancho, M.A.H. Vozmediano,
New J. Phys. {\bf 11}, 095017 (2009).

\bibitem{castro3}
E.V. Castro, M.P. L\'opez-Sancho, M.A.H. Vozmediano,
Phys. Rev. Lett. {\bf 104}, 036802 (2010).

\bibitem{castro4}
E.V. Castro, M.P. L\'opez-Sancho, M.A.H. Vozmediano,
Phys. Rev. B {\bf 84}, 075432 (2011).





\bibitem{cresti}
A. Cresti, N. Nemec, B. Biel, G. Niebler, F. Triozon,
G. Cuniberti, and S. Roche,
Nano Res. {\bf 1}, 361 (2008).


\bibitem{huang}
Y.C. Huang, C.P. Chang, W.S. Su, and M.F. Lin,
J. Appl. Phys. {\bf 106}, 013711 (2009).

\bibitem{klymenko}
Y. Klymenko and O. Shevtsov,
Eur. Phys. J. B {\bf 69}, 383 (2009).



\bibitem{LdS}
J.M.B. Lopes dos Santos, N.M.R. Peres, and A.H. Castro Neto,
Phys. Rev. Lett. {\bf 99}, 256802 (2007).


\bibitem{mele1}
E.J. Mele,
Phys. Rev. B {\bf 81}, 161405 (2010).

\bibitem{mele2}
E.J. Mele,
J. Phys. D: Appl. Phys. {\bf 45}, 154004 (2012).

\bibitem{shallcross}
S. Shallcross, S. Sharma, E. Kandelaki, and O.A. Pankratov,
Phys. Rev. B {\bf 81}, 165105 (2010).

\bibitem{morell}
E. Su\'arez Morell, J.D. Correa, P. Vargas, M. Pacheco, and
Z. Barticevic,
Phys. Rev. B {\bf 82}, 121407 (2010).

\bibitem{degail}
R. de Gail, M.O. Goerbig, F. Guinea,
G. Montambaux, and A. H. Castro Neto,
Phys. Rev. B {\bf 84}, 045436 (2011).

\bibitem{mac}
R. Bistritzer and A.H. MacDonald,
Proc. Nat. Ac. Sc. {\bf 108}, 12233 (2011).

\bibitem{morell2}
E. Su\'arez Morell, P. Vargas, L. Chico, and L. Brey,
Phys. Rev. B {\bf 84}, 195421 (2011).

\bibitem{moon1}
P. Moon and M. Koshino,
Phys. Rev. B {\bf 85}, 195458 (2012).

\bibitem{moon2}
P. Moon and M. Koshino,
arXiv:1302.5218 (2013).


\bibitem{sanjose}
P. San-Jos\'e, J. Gonz\'alez, and F. Guinea,
Phys. Rev. Lett. {\bf 108}, 216802 (2012).

\bibitem{tabert}
C. J. Tabert and E. J. Nicol,
Phys. Rev. B {\bf 87}, 121402 (2013).


\bibitem{kaasbjerg}
K. Kaasbjerg, K.S. Thygesen, and K.W. Jacobsen,
Phys. Rev. B {\bf 85}, 115317 (2012).

\bibitem{kaasbjerg2}
K. Kaasbjerg, A.-P. Jauho, and K.S. Thygesen,
arXiv:1206.2003 (2012).

\bibitem{feng2}
W. Feng, Y. Yao, W. Zhu, J. Zhou, W. Yao, and D. Xiao,
Phys. Rev. B {\bf 86}, 165108 (2012).

\bibitem{kadantsev}
E.S. Kadantsev and P. Hawrylak,
Solid State Comm. {\bf 152}, 909 (2012).

\bibitem{kosmider}
K. Ko\'smider and J. Fern\'andez-Rossier,
Phys. Rev. B {\bf 87}, 075451 (2013).

\bibitem{zibouche}
N. Zibouche, A. Kuc, and T. Heine,
arXiv:1302.3478 (2013).


\bibitem{li4}
X. Li, J.T. Mullen, Z. Jin, K.M. Borysenko, M. Buongiorno Nardelli, K.W. Kim,
arXiv:1301.7709 (2013).

\bibitem{bromley}
R.A. Bromley, R.B. Murray, and A.D. Yoffe,
J. Phys. C: Solid State Phys. {\bf 5}, 759 (1972).

\bibitem{korma}
A. Kormanyos, V. Zolyomi, N.D. Drummond, P. Rakyta,
G. Burkard, and V.I. Fal'ko,
arXiv:1304.4084 (2013).


\bibitem{asgari}
H. Rostami, A.G. Moghaddam, and R. Asgari,
arXiv:1302.5901 (2013).

\bibitem{zahid}
F. Zahid, L. Liu, Y. Zhu, J. Wang, and H. Guo,
arXiv:1304.0074 (2013).


\bibitem{sl}
J.C. Slater and G.F. Koster,
Phys. Rev. {\bf 94}, 1498 (1954).



\bibitem{ceperley_alder_1980}
D.M. Ceperley, and B.J. Alder,
Phys. Rev. Lett. {\bf 45}, 566 (1980).

\bibitem{perdew}
J. Perdew and A. Zunger, Phys. Rev. B {\bf 23}, 5048 (1981)

\bibitem{arsan99}
E. Artacho, D. S\'anchez-Portal, P. Ordej\'on,
A. Garc\'{\i}a and J.M. Soler,
Phys. Stat. Sol. (b) {\bf 215}, 809 (1999).

\bibitem{doran}
N.J. Doran, B. Ricco, D.J. Titterington, and G. Wexler,
J. Phys. C: Sol. State Phys. {\bf 11}, 685 (1978).

\bibitem{noteTB}
A good fitting agreement with DFT data was
shown in Ref. \onlinecite{zahid},
but using a larger, non-orthogonal basis set, and involving
up to 96 fitting parameters.

\bibitem{numrecipes}
W. H. Press, S. A. Teukolsky, W. T. Vetterling 
and B. P. Flannery, {\em Numerical Recipes} 
(Cambridge University Press, 1992).
  

\end{thebibliography}
\end{document}